\documentclass[aps, prb, a4paper,twocolumn,superscriptaddress]{revtex4-1}
\usepackage{graphicx}
\usepackage{amsmath}
\usepackage{amssymb}
\usepackage{bbold}
\usepackage{braket}
\usepackage{multirow}
\usepackage{natbib}
\graphicspath{{graphics/}}
\begin{document}

\title{A solvable quantum model of dynamic nuclear polarization in optically driven quantum dots}
\author{Thomas Nutz}
\email{nutzat@gmail.com}
\affiliation{Controlled Quantum Dynamics Theory Group, Imperial College London, London SW7 2AZ, United Kingdom}
\author{Edwin Barnes}
\affiliation{Department of Physics, Virginia Tech, Blacksburg, Virginia 24061, USA}
\author{Sophia E. Economou}
\affiliation{Department of Physics, Virginia Tech, Blacksburg, Virginia 24061, USA}
\date{\today}

\begin{abstract}
We present a quantum mechanical theory of optically induced dynamic nuclear polarization applicable to quantum dots and other interacting spin systems. The exact steady state of the optically driven coupled electron-nuclear system is calculated under the assumption of uniform hyperfine coupling strengths (box model) for an arbitrary number of nuclear spins. This  steady state is given by a tractable expression that allows for an intuitive interpretation in terms of spin-flip rates. Based on this result, we investigate the nuclear spin behaviour for different experimental parameter regimes and find that our model reproduces the flat-top and triangular absorption line shapes seen in various experiments (line dragging) under the assumption of fast electron spin dephasing due to phonons and co-tunneling. The mechanism responsible for line dragging has been a matter of controversy so far; in contrast with previous works, we show that the effect can be explained solely in terms of the contact hyperfine interaction, without the need to introduce non-collinear terms or other types of electron-nuclear interactions. Furthermore we predict a novel nuclear spin polarization effect: under particular, experimentally realistic conditions, the nuclear spin system tends to become sharply polarized in such a way as to cancel the effect of the external magnetic field. This effect can therefore suppress electron spin dephasing due to inhomogeneous broadening, which could have important repercussions for quantum technological applications.
\end{abstract}

\maketitle

\section{Introduction}

The coupling between an optically driven electron and the $\sim 10^5$ nuclear spins inside a semiconductor quantum dot gives rise to drastic and unexpected phenomena that have been challenging to explain in a consistent way. While quantum mechanical models of nuclear spin dynamics in pulsed laser experiments \cite{barnes_electron-nuclear_2011, economou_theory_2014, beugeling_quantum_2016, beugeling_influence_2017} have been used to explain frequency focusing effects \cite{greilich_nuclei-induced_2007, carter_directing_2009}, the underlying assumption of optical and nuclear processes taking place on different timescales does not hold for experiments with continuous wave (CW) laser drive. In fact a consistent quantum mechanical description with predictive power for important CW effects such as line dragging \cite{latta_confluence_2009, hogele_dynamic_2012, xu_optically_2009, ladd_pulsed_2010} and nuclear polarization induced by quasiresonant excitation \cite{braun_bistability_2006, eble_dynamic_2006, krebs_hyperfine_2008, lai_knight-field-enabled_2006, urbaszek_nuclear_2013} is still lacking. Such a theory of optically induced dynamic nuclear polarization (DNP) could be crucial for quantum dot (QD) based technologies to realize their immense potential. QD single-photon sources have recently achieved unprecedented brightness \cite{wang_toward_2018} and near-perfect single-photon purity \cite{hanschke_quantum_2018, somaschi_near-optimal_2016, ding_-demand_2016}, as well as generated entangled states of several photons \cite{schwartz_deterministic_2016-1}. For these systems to become fundamental building blocks of a quantum computer, however, the coherence of a trapped electron spin needs to be preserved in the presence of the nuclear spin environment. DNP can be used to strongly reduce the variance of the effective magnetic field acting on the electron spin \cite{burkard_coupled_1999}, which greatly increases the decoherence time of this qubit system. Understanding and controlling nuclear spins therefore constitutes a crucial step in advancing quantum dot science and technology.

We present a model of DNP induced by CW lasers that features two novel, critical elements in a DNP theory. First, our model allows us to capture the full quantum state of thousands of nuclear spins interacting with an electron spin in a QD. This approach makes it possible to consider the collective quantum mechanical behaviour of the nuclear spin system and leads to a tractable solution in the so-called ``box model'' limit, in which the electron-nuclear contact hyperfine couplings are taken to be uniform. The box model has been shown to successfully capture the DNP-driven frequency focusing effect \cite{barnes_electron-nuclear_2011, economou_theory_2014} as well as short-time free induction decay \cite{barnes_master_2011, barnes_nonperturbative_2012}. More importantly, we simultaneously take into account optical driving, decay, and electron-nuclear interactions. The interplay of these mechanisms turns out to be centrally important to understanding DNP in these systems, as energy conserving second-order processes emerge that transfer spin polarization between electron and nuclear systems even in the presence of a large external magnetic field.
Our model can be viewed as minimal in the sense that it includes only the contact hyperfine interaction, the electron Zeeman interaction, and optical driving. Additional interactions such as quadrupolar hyperfine terms or nuclear dipole-dipole interactions are not included. Our work thus sheds light on which DNP effects can be explained solely in terms of the contact interaction.

\begin{figure}
\includegraphics[width = \linewidth]{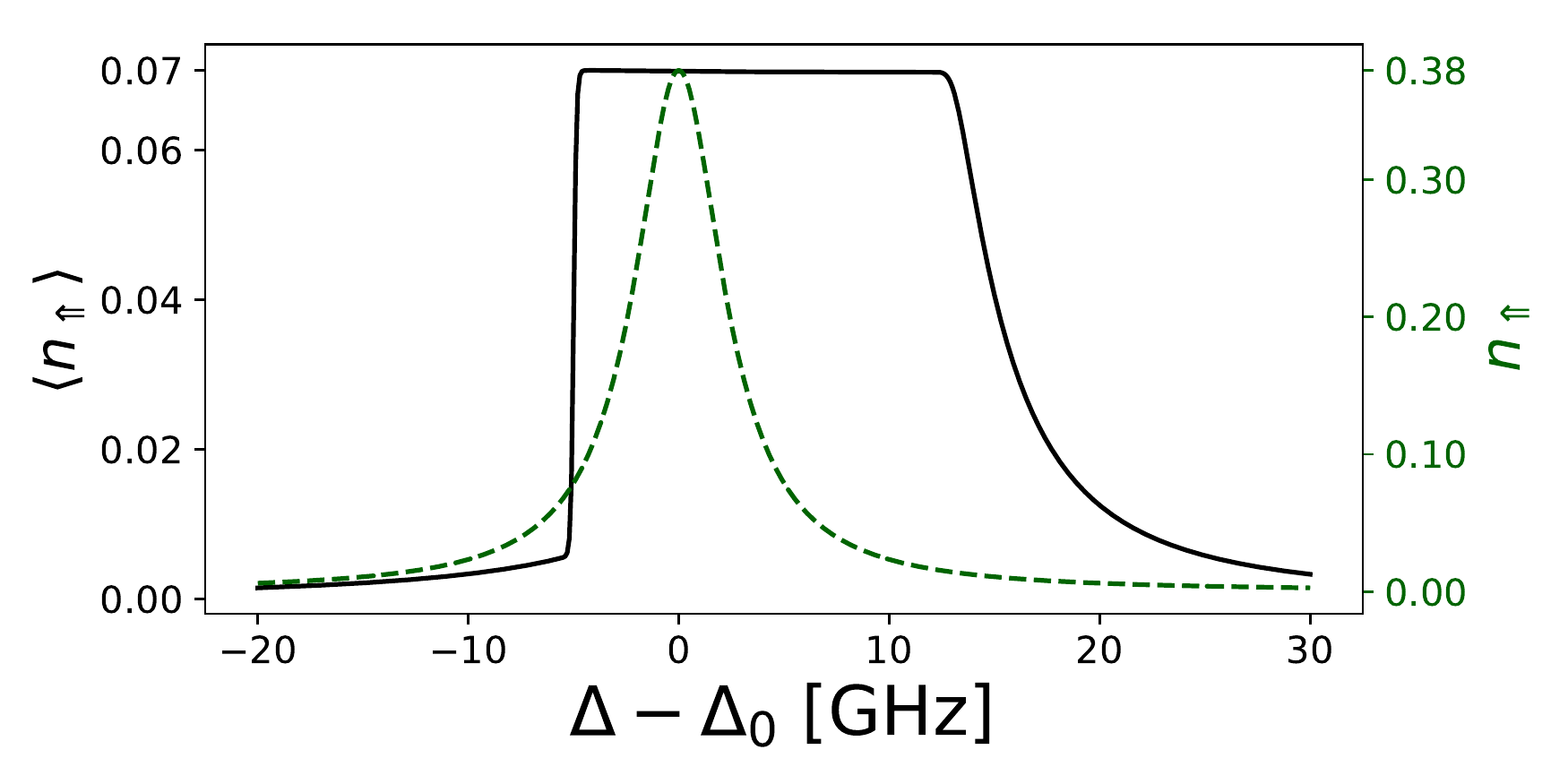}
\caption{Trion population averaged over steady-state nuclear spin polarizations (black).
The nuclear spin system adjusts in such a way that the optical transition remains at a fixed detuning over a range of laser frequencies parametrized by $\Delta$. The green dashed line shows the Lorentzian line shape of the driven $\ket{\uparrow} \leftrightarrow \ket{\Uparrow}$ transition in the absence of nuclear spins. Parameters in GHz: $\Gamma = 1$, $\gamma = 0.01$, $A = 0.01$, $\Omega = 1$, $\omega_e = -150$, $J = 2000$, $\kappa_{\uparrow} = 0.01$, $\kappa_{\downarrow} = 0.011$, $\eta = 1.5$, $\Delta_0 = \omega_e/2$.
}
\label{fig:line shape}
\end{figure}

We find an exact and analytically tractable expression for the steady state of the optically driven, coupled electron-nuclear spin system for an arbitrary number of nuclear spins in the box model approximation. This exact expression allows for an intuitive interpretation in terms of spin-flip rates and describes a rich variety of DNP effects. In particular we find that our model reproduces both the flat-top and the triangular absorption line shapes seen in experiments \cite{latta_confluence_2009, hogele_dynamic_2012} and referred to as line dragging under the assumption of a strong electron spin dephasing mechanism on top of the electron-nuclear interaction.  An example of a flat-top line shape is shown in Fig. \ref{fig:line shape}. Furthermore we find an interplay of second-order flip-flop processes leading to sharply peaked nuclear spin polarization probability distributions (NSPPDs). This novel DNP effect can give rise to nuclear spin polarization that tends to cancel the effect of an external magnetic field and thereby leads to degenerate electronic transitions. We refer to this effect as ``Zeeman suppression''. Due to the strongly reduced variance of the NSPPD, this effect could greatly reduce electron spin dephasing due to inhomogeneous broadening, which along with the degenerate transitions established by the effect is a crucial requirement for a QD source of entangled photons as proposed in Ref. \onlinecite{lindner_proposal_2009}.

Our exact solution for the steady state of an optically driven, coupled electron-nuclear spin system provides an explanation for experimentally observed DNP effects and predicts interesting novel phenomena, which could pave the way towards QD-based quantum technologies. As an exact solution of a mesoscopic quantum mechanical problem, our result can furthermore serve as a touchstone for approximate methods going beyond the box model approximation. 

This paper begins with the introduction of the model in section II. The steady-state solution of this model is derived in section III and used in sections IV and V to investigate the DNP effects of line dragging and Zeeman suppression, respectively. An overview of the effects found is shown in Tab. \ref{tab:effects} and the experimental parameters are summarized in Tab. \ref{tab:parameters}.

\begin{table}[h]
\begin{tabular}{| l | l | l | p{2.9cm} |}
  \hline
  \textbf{Transition} & \textbf{Detuning} & \textbf{Dephasing} & \textbf{Effect} \\
  \hline
  blue ($\omega_e < 0$) & \multirow{2}{1.4cm}{$\Delta \approx \frac{\omega_e}{2}$ } & \multirow{2}{1.5cm}{fast ($\eta_t \gtrsim \Gamma$)} & line dragging \\
  \cline{1-1}
  \cline{4-4}
  red ($\omega_e > 0$) & & & anti-dragging \\
  \hline  
  \multirow{2}{1.8cm}{blue or red} & $\Delta < 0$ & \multirow{2}{1.4cm}{slow ($\eta_t \ll \Gamma$)} & Zeeman suppression \\
  \cline{2-2}
  \cline{4-4}
   & $\Delta > 0$ &  & anti-dragging \\
  \hline
\end{tabular}
\caption{DNP effects found in four different parameter regimes when a right-circularly polarized CW laser is applied along the growth axis of the QD. In this case, the laser excites the electron state $\ket{\uparrow}$ up to the trion state $\ket{\Uparrow}$. $\omega_e$ is the Zeeman splitting between the electronic spin states, and $\Delta=\epsilon_\Uparrow-\omega_{drive}$ is the laser detuning relative to the midway point between the two electron spin states (taken to be the zero of energy). Because the electronic $g$-factor is negative, $\omega_e<0$ means that $\ket{\uparrow}$ has a lower Zeeman energy than $\ket{\downarrow}$, and thus the $\ket{\uparrow}\leftrightarrow \ket{\Uparrow}$ transition has larger frequency (blue).}
\label{tab:effects}
\end{table}

\section{The optically driven box model}

We consider a state $\rho$ of the electron-nuclear system subject to a Lindblad equation
\begin{equation}
\dot{\rho} = -\frac{i}{\hbar}[H,\rho] + \sum_i \left( L_i \rho L_i^{\dagger} - \frac{1}{2}(L_i^{\dagger}L_i \rho + \rho L_i^{\dagger}L_i) \right),
 \label{eq:Lindblad}
\end{equation}
where $i \in \{ \Gamma, \gamma, \uparrow, \downarrow, \eta_1, \eta_2 \}$ with the Lindblad operators $L_i$ defined below. The Hamiltonian $H = H_{\mathrm{el}} + H_{\mathrm{hf}}$ comprises a purely electronic term $H_{\mathrm{el}}$ describing the electron Zeeman effect and driving by a classical optical field as well as the hyperfine interaction $H_{\mathrm{hf}}$ which couples the electron to the nuclear spin system. The electronic component is given in the frame rotating at the driving frequency in the rotating wave approximation by
\begin{equation}
  H_{\mathrm{el}} = \Omega (\ket{\Uparrow}\bra{\uparrow} + \ket{\uparrow}\bra{\Uparrow}) +  \Delta \ket{\Uparrow}\bra{\Uparrow} + \frac{\omega_e}{2}(\ket{\uparrow}\bra{\uparrow} - \ket{\downarrow}\bra{\downarrow}),
  \label{eq: H_el}
\end{equation}
where $\Omega$ is the Rabi frequency of the drive, $\Delta$ the difference between trion energy (including the hole Zeeman contribution) and driving frequency, and $\omega_e$ the electronic Zeeman splitting. Note that we do not include the trion state $\ket{\Downarrow}$ in our model.  This omission is justified for experiments where the laser drives only one of the two optically active transitions $\ket{\uparrow} \leftrightarrow \ket{\Uparrow}$ and $\ket{\downarrow} \leftrightarrow \ket{\Downarrow}$ due to its polarization or detuning from one transition in a large external magnetic field. The electron-nuclear coupling is given by the contact hyperfine Hamiltonian, which in the box model approximation \cite{kozlov_exactly_2007} takes the form 
\begin{equation}
  H_{\mathrm{hf}}= A(S_zI_z + \frac{1}{2}(S^+I^- + S^-I^+)),
  \label{eq: Box model}
\end{equation}
where $S_z = 1/2\ (\ket{\uparrow}\bra{\uparrow} - \ket{\downarrow}\bra{\downarrow})$, $I_z = \sum_i I_z^i$ and accordingly for $I^{+/-}$ and $S^{+/-}$. 

We furthermore include spontaneous decay of the trion state $\ket{\Uparrow}$ via Lindblad terms $L_{\Gamma} = \sqrt{\Gamma} \ket{\uparrow}\bra{\Uparrow}$, $L_{\gamma} = \sqrt{\gamma} \ket{\downarrow}\bra{\Uparrow}$, where $\Gamma \gg \gamma$ are the vertical and diagonal decay rates, respectively. Electron spin relaxation is described by $L_{\uparrow} = \sqrt{\kappa_{\uparrow}} \ket{\downarrow}\bra{\uparrow}$, $L_{\downarrow} = \sqrt{\kappa_{\downarrow}} \ket{\uparrow}\bra{\downarrow}$, and electron spin dephasing by $L_{\eta 1} = \sqrt{\eta} \ket{\downarrow}\bra{\downarrow}$, $L_{\eta 2} = \sqrt{\eta} \ket{\uparrow}\bra{\uparrow}$. Electron spin relaxation occurs mainly due to co-tunneling of electrons in and out of the dot \cite{dreiser_optical_2008}, where the two different co-tunneling rates can be chosen consistently with a thermal electron spin state in the absence of optical driving, i.e. $\kappa_{\downarrow}/\kappa_{\uparrow} = \exp \big(- \frac{\omega_e}{k_BT}\big)$. Electron spin dephasing on top of nuclear dephasing can be attributed to phonons \cite{grosse_phonons_2008}.

\begin{figure}
\includegraphics[width=\linewidth]{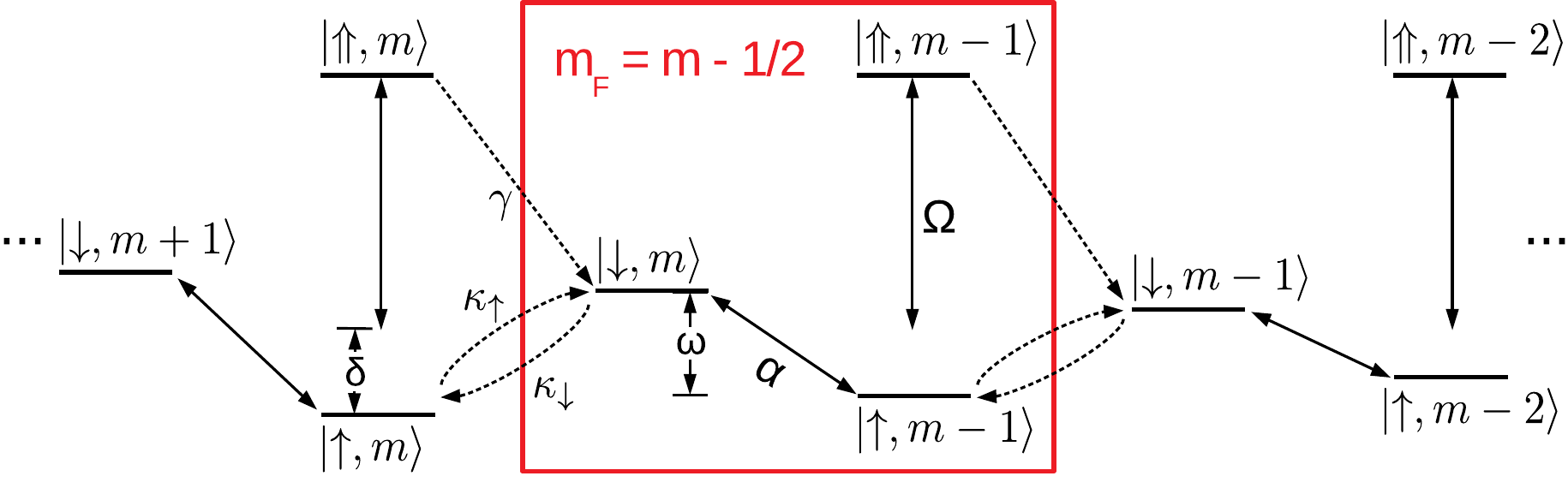}
\caption{The box model Hamiltonian is block diagonal in blocks corresponding to eigenvalues of $F_z = I_z + S_z$. The level scheme inside the red box represents such a subspace referred to as an $F$ block. Within each $F$ block, states are coupled via optical driving and the hyperfine interaction, represented by double sided arrows labelled $\Omega$ and $\alpha$, respectively. Processes that transfer population between $F$ blocks are co-tunneling at rates $\kappa_{\uparrow}$ and $\kappa_{\downarrow}$ as well as diagonal decay at rate $\gamma$.}
 \label{fig:box chain}
\end{figure}

The box model approximation greatly simplifies the problem, as an electron-nuclear spin state $\ket{\downarrow}\otimes \ket{m}$, where $I_z \ket{m} = m \ket{m}$, only couples to $\ket{\uparrow}\otimes \ket{m-1}$ and vice versa. The hyperfine Hamiltonian is therefore block diagonal in blocks corresponding to eigenvalues $m_F$ of the total electron-nuclear spin operator $F_z = S_z + I_z$. This block diagonal structure is illustrated in Fig. \ref{fig:box chain}.

\section{The steady-state solution}

We want to calculate the steady state of the system, i.e. the state $\rho_{\mathrm{std}}$ that satisfies $\dot{\rho} = 0$ with $\dot{\rho}$ given by Eq. \eqref{eq:Lindblad}. We first note that due to the block diagonality of $H$ all coherences between states of different $m_F$ vanish. The remaining linear system can be solved by a decoupling procedure. We consider the steady-state condition for the trace of a $m_F = m-1/2$ block within $\dot{\rho} = 0$:
\begin{equation}
\begin{split}
& \dot{\rho}_{\downarrow}(m)+\dot{\rho}_{\uparrow}(m-1)+\dot{\rho}_{\Uparrow}(m-1) = \\
& \gamma (\rho_{\Uparrow}(m)- \rho_{\Uparrow}(m-1))+ \\
& \kappa_{\downarrow} \left(\rho_{\downarrow}(m-1) - \rho_{\downarrow}(m)\right) + \kappa_{\uparrow} \left( \rho_{\uparrow}(m) - \rho_{\uparrow}(m-1) \right) \\
&= 0,
\end{split}
 \label{eq:test}
\end{equation}
where we use the notation $\rho_{e}(m) = \braket{m,e | \rho | m, e}$ with $e \in \{ \downarrow, \uparrow, \Uparrow \}$ for state populations. Adding such trace equations for adjacent $m_F$ blocks starting with the trace equation for the $m_F = J+1/2$ block,
\begin{equation}
\gamma \rho_{\Uparrow}(J) + \kappa_{\uparrow} \rho_{\uparrow}(J) = \kappa_{\downarrow} \rho_{\downarrow}(J),
\label{eq:J block trace} 
\end{equation}
one arrives at the series of equations
\begin{equation}
\gamma \rho_{\Uparrow}(m) + \kappa_{\uparrow} \rho_{\uparrow}(m) = \kappa_{\downarrow} \rho_{\downarrow}(m)\ \ \textrm{for}\ \ J \geq m \geq -J.
 \label{eq:relative normalization equations}
\end{equation}
The steady-state equations 
\begin{equation}\begin{split}
\dot{\rho}_{\uparrow}(m) = & \braket{\uparrow, m | [H,\rho] | \uparrow, m } + \\ 
& \Gamma \rho_{\Uparrow}(m) - \kappa_{\uparrow} \rho_{\uparrow}(m) + \kappa_{\downarrow} \rho_{\downarrow}(m) = 0, \\
\dot{\rho}_{\downarrow}(m) = & \braket{\downarrow, m | [H,\rho] | \downarrow, m } + \\ 
& \gamma \rho_{\Uparrow}(m) +\kappa_{\uparrow} \rho_{\uparrow}(m) - \kappa_{\downarrow} \rho_{\downarrow}(m) = 0,
\end{split} \label{eq:rho dot Neumann and Lindblad bits}
\end{equation}
simplify to
\begin{equation}\begin{split}
\dot{\rho}_{\uparrow}(m) = & \braket{\uparrow, m | [H,\rho] | \uparrow, m } + 
 (\Gamma + \gamma ) \rho_{\Uparrow}(m) = 0, \\
\dot{\rho}_{\downarrow}(m) = & \braket{\downarrow, m | [H,\rho] | \downarrow, m } = 0
\end{split} \label{eq:rho dot Neumann and Lindblad bits 2}
\end{equation}
using Eq. \eqref{eq:relative normalization equations}. The resulting set of equations for any $m_F = m-1/2$ block is therefore identical to the equations generated by a steady-state Lindblad equation $\mathcal{L} \rho_{m_F} = 0$ with Hamiltonian $H_{m_F} = P_{m_F} H P_{m_F}$ and Lindblad operators $L_{v} = \sqrt{\Gamma + \gamma} \ket{\uparrow}\bra{\Uparrow}$, $L_{\uparrow} = \sqrt{\kappa_{\uparrow} + \eta} \ket{\uparrow}\bra{\uparrow}$, $L_{\downarrow} = \sqrt{\kappa_{\downarrow} + \eta} \ket{\downarrow}\bra{\downarrow}$. Here $P_{m_F}$ is the projector onto the $m_F$ subspace spanned by $\ket{\downarrow , m}$, $\ket{\uparrow , m-1}$, $\ket{\Uparrow , m-1}$.

The steady-state equation $\dot{\rho} = 0$ has now been decoupled into a $3 \times 3$ set of steady-state equations $\mathcal{L} \rho_{m_F} = 0$, referred to as the ``one-block problem'', and a set of equations given by Eq. \eqref{eq:relative normalization equations} relating these one-block problems, which we refer to as the ``all-blocks problem''. While the one-block problem involves finding the steady state of a Lindblad equation in a three-dimensional state space, the all-blocks problem is solved by normalizing these one-block solutions in such a way as to satisfy Eq. \eqref{eq:relative normalization equations}. In what follows, we first show how to perform this normalization procedure assuming the single-block problem has been solved. We then show how to solve the single-block problem to obtain an explicit expression for the full steady-state density matrix. 

The all-blocks problem can be formulated in terms of rate equations in the probabilities
\begin{equation}
p_{m_F} = \rho_{\downarrow}(m) + \rho_{\uparrow}(m-1) + \rho_{\Uparrow}(m-1),
 \label{eq:mF probabilities definition}
\end{equation}
where $m = m_F - 1/2$. We write 
\begin{equation}\begin{split}
\dot{p}_{m_F}(t) =  & - \left( r_{\downarrow}^{m_F}(t)+ r_{\uparrow}^{m_F}(t) \right)  p_{m_F}(t) + \\
& r_{\downarrow}^{m_F + 1}(t)\ p_{m_F + 1}(t) + r_{\uparrow}^{m_F - 1}(t)\ p_{m_F - 1}(t),
\end{split}
 \label{eq:rate equation}
\end{equation}
where $r_{\downarrow / \uparrow}^{m_F}(t)$ denotes the rate of the system to transition from a state with total electron-nuclear spin $m_F$ downwards (upwards) to a state of $m_F = m_F - 1$ ($m_F = m_F + 1$) at time $t$. These rates can be identified as 
\begin{equation}\begin{split}
r_{\downarrow}^{m_F}(t) &= \kappa_{\uparrow} n_{\uparrow}(m_F,t) + \gamma n_{\Uparrow}(m_F,t), \\
r_{\uparrow}^{m_F}(t) &= \kappa_{\downarrow} n_{\downarrow}(m_F,t),
\end{split}
 \label{eq:rates i.t.o. dmat els}
\end{equation}
where $n_e(m_F,t)$ are the diagonal elements of a normalized $3 \times 3$ density matrix $\rho_{m_F}(t)$ obeying $\braket{e | \rho(t) | e} = n_e(m_F,t)\ p_{m_F}(t)$ and $\sum_e n_{e}(m_F) = 1$ with $e \in \{\Uparrow, \uparrow, \downarrow \}$.

The steady state is given by $\dot{p}_{m_F} = 0$. Using Eq. \eqref{eq:relative normalization equations}, which also holds in the steady state, one finds that 
\begin{equation}
\frac{p_{m_F}}{p_{m_F-1}} = \frac{r_{\uparrow}^{m_F-1}}{r_{\downarrow}^{m_F}},
 \label{eq:rate pic  steady-state}
\end{equation}
where dropping the time dependence indicates steady-state rates and probabilities. This equation can be interpreted as a statement of zero probability flow across each side of the $m_F$ box shown in Fig. \ref{fig:box chain}. Given the steady-state rates $r_{\downarrow \uparrow}^{m_F}$, which can be determined by solving the one-block problem, the all-blocks problem is solved by
\begin{equation}\begin{split}
p_{m_F} &= \frac{1}{N} \frac{1}{r_{\downarrow}^{m_F}} \prod_{i=1}^{m_F - 1} \left( \frac{r_{\uparrow}^i}{r_{\downarrow}^i} \right) \ \ \ \mathrm{for}\ m_F>0, \\
p_{m_F} &= \frac{1}{N} \frac{1}{r_{\uparrow}^{m_F}} \prod_{i=0}^{m_F + 1} \left( \frac{r_{\downarrow}^i}{r_{\uparrow}^i} \right) \ \ \ \mathrm{for}\ m_F < 0, \\
p_{0} &= \frac{1}{N} \frac{1}{r_{\uparrow}^{0}},
\end{split} \label{eq:all-blocks solution}
\end{equation}
where $N$ is an overall normalization constant. 

The  steady state of the optically driven box model is therefore found by solving the one-block problem $\mathcal{L} \rho_{m_F} = 0$ to obtain the three relative populations $n_{\downarrow}(m_F)$, $n_{\uparrow}(m_F)$, and $n_{\Uparrow}(m_F)$, calculating the spin-flip rates given by Eq. \eqref{eq:rates i.t.o. dmat els}, and using these rates to calculate the polarization probability distribution given in Eq. \eqref{eq:all-blocks solution}. Eq. \eqref{eq:rate pic  steady-state} and Eq. \eqref{eq:all-blocks solution} have an intuitive graphical interpretation that can directly lead to a qualitative understanding of the NSPPD $p_{m_F}$ based on properties of the one-block solution, as shown in Fig. \ref{fig:line dragging}.

We now return to the one-block problem and show how to solve the  $3\times 3$ set of equations to obtain an explicit expression for the steady state. To compactify the notation in what follows, we introduce the auxiliary parameters $\alpha = \frac{A}{2}\sqrt{J(J+1) - m(m-1)}$, $\omega = -\omega_e - A m + A/2$, $\delta = \Delta - \left( \omega_e + A (m - 1) \right)/2 $, $\Gamma^{\prime} = \Gamma + \gamma$, $\eta_{\uparrow} = \eta + \kappa_{\uparrow}$, and $\eta_{\downarrow} = \eta + \kappa_{\downarrow}$. Note that $\delta$ is the detuning of the laser from the optical transition $\ket{\uparrow,m}\leftrightarrow \ket{\Uparrow,m}$. We also note that in the one-block problem, the co-tunneling rates $\kappa_{\uparrow / \downarrow}$ appear as part of dephasing rather than as relaxation rates, because the effect of electron spin relaxation in the one-block problem cancels out when using Eq. \eqref{eq:relative normalization equations}. We therefore refer to $\eta_{\uparrow / \downarrow}$ as combined dephasing rates, keeping in mind that this dephasing is partially a consequence of relaxation. The relative populations $n_e(m_F = m - 1/2)$ for $e \in {\downarrow, \uparrow, \Uparrow}$ are then found as the steady state of the Lindblad equation with Hamiltonian
\begin{equation}\begin{split}
H =  \omega &\ket{\downarrow}\bra{\downarrow} + \delta \ket{\Uparrow}\bra{\Uparrow} \\
 + \alpha \big( &\ket{\downarrow}\bra{\uparrow} + \ket{\uparrow}\bra{\downarrow} \big) + \Omega \big(\ket{\uparrow}\bra{\Uparrow} + \ket{\Uparrow}\bra{\uparrow}\big),
\end{split}
 \label{eq:Hamiltonian aux symbols}
\end{equation}
where the zero of energy has been reset by subtracting $\omega_e \mathbb{1} /2$ w.r.t. Eq. \ref{eq: H_el},
and Lindblad operators $L_{\Gamma^{\prime}} = \sqrt{\Gamma^{\prime}}\ket{\uparrow}\bra{\Uparrow}$, $L_{\uparrow} = \sqrt{\eta_{\uparrow}}\ket{\uparrow}\bra{\uparrow}$, $L_{\downarrow} = \sqrt{\eta_{\downarrow}}\ket{\downarrow}\bra{\downarrow}$.  We find

\begin{equation}\begin{split}
n_{\Uparrow} &= \frac{D}{D(2T+1) + 2 \left(\frac{\alpha}{\Omega} \right)^2 N_1 - N_2}, \\
n_{\uparrow} &= \frac{TD+\left(\frac{\alpha}{\Omega} \right)^2 N_1}{D(2T+1) + 2 \left(\frac{\alpha}{\Omega} \right)^2 N_1 - N_2},\\
n_{\downarrow} &= \frac{TD+\left(\frac{\alpha}{\Omega} \right)^2 N_1 - N_2}{D(2T+1) + 2 \left(\frac{\alpha}{\Omega} \right)^2 N_1 - N_2},
\end{split} \label{eq: steady-state populations} \end{equation}

where

\begin{equation}\begin{split} T &= 1 + \frac{\Gamma^{\prime} \Gamma^{\prime}_{\uparrow}}{4 \Omega^2} + \frac{\Gamma^{\prime} \delta^2}{\Gamma^{\prime}_{\uparrow} \Omega^2},\\
D &= \frac{1}{\Gamma^{\prime}}\left( \frac{\Gamma^{\prime}_{\downarrow}}{4} + \frac{\alpha^2}{\Gamma^{\prime}_{\uparrow}}+ \frac{\Omega^2}{\eta_t}+ \frac{(\delta - \omega)^2}{\Gamma^{\prime}_{\downarrow}} \right), \\
N_1 &= \frac{1}{4}  + \frac{\Omega^2}{\eta_{t}\Gamma^{\prime}_{\downarrow}}+ \frac{2 \delta (\omega - \delta) + \alpha^2}{\Gamma^{\prime}_{\uparrow}\Gamma^{\prime}_{\downarrow}} - \frac{\delta^2}{(\Gamma^{\prime}_{\uparrow})^2}, \\
N_2 &= \frac{1}{4} + \frac{\Omega^2}{\eta_{t}\Gamma^{\prime}_{\downarrow}} + \frac{\delta (\omega - \delta) + \alpha^2}{\Gamma^{\prime}_{\uparrow}\Gamma^{\prime}_{\downarrow}} + \frac{\omega}{\eta_{t}}\left( \frac{\delta - \omega}{\Gamma^{\prime}_{\downarrow}} + \frac{\delta}{\Gamma^{\prime}_{\uparrow}} \right),
\end{split}
 \label{eq:auxiliary functions}
\end{equation}
and we use the short forms $\Gamma^{\prime}_{\uparrow / \downarrow} = \Gamma^{\prime} + \eta_{\uparrow / \downarrow}$ and $\eta_{t} = \eta_{\downarrow} + \eta_{\uparrow}$. This solution of the one-block problem lets us calculate the spin-flip rates $r^{m_f}_{\uparrow / \downarrow}$ following Eq. \eqref{eq:rates i.t.o. dmat els}, which in turn specify the NSPPD $p_{m_F}$. The density matrix $\rho_{\mathrm{std}}$ with populations given by $\braket{e | \rho_{\mathrm{std}} | e} = n_e(m_F) p_{m_F}$ is the exact steady state of the optically driven box model specified by the Lindblad equation in Eq. \eqref{eq:Lindblad}. This steady-state solution is our main result and will be used in the following to explore interesting DNP effects such as line dragging and Zeeman suppression.

\begin{table}
\begin{tabular}{| p{1.3cm} | p{7cm} |}
  \hline
  \textbf{Symbol} & \textbf{Meaning}  \\
  \hline
  $\Gamma$ & vertical decay rate  ($\ket{\Uparrow} \rightarrow \ket{\uparrow}$) \\
  $\gamma$ & diagonal decay rate  ($\ket{\Uparrow} \rightarrow \ket{\downarrow}$)\\
  $A$ & hyperfine coupling constant \\
  $ J $ & total nuclear spin quantum number \\
  $\Omega$ & Rabi frequency of the $\ket{\Uparrow} \leftrightarrow \ket{\uparrow}$ driving \\
  $\Delta$ & laser drive detuning $\epsilon_{\Uparrow} -  \omega_{\mathrm{drive}}$ (does not include Zeeman splitting) \\
  $\omega_e$ & Zeeman splitting of the electron spin states \\
  $\eta$ & electron spin dephasing rate \\
  $\kappa_{\uparrow}$ ($\kappa_{\downarrow}$) & co-tunneling rate from up to down (down to up) \\
  $\eta_{\uparrow}$ ($\eta_{\downarrow}$) & combined dephasing rate $\eta + \kappa_{\uparrow}$ ($\eta + \kappa_{\downarrow}$)\\
  $\omega$ & total splitting of states $\ket{\downarrow, m}$ and $\ket{\uparrow, m-1}$: \newline
  $\omega = \epsilon_{\downarrow,m}-\epsilon_{\uparrow,m-1} = -\omega_e - A\ m + A/2$ \\
  $\delta$ & total detuning (includes Zeeman and hyperfine splitting): $\delta =\Delta - (\omega_e + A(m-1))/2$  \\
  $\alpha$ & effective hyperfine coupling: \newline
  $\alpha = \frac{A}{2}\sqrt{J(J+1) - m(m-1)}$ \\
  \hline
\end{tabular}
\caption{Parameters of the optically driven box model.}
\label{tab:parameters}
\end{table}

\section{Line dragging}
The one-block solution (Eq. \eqref{eq: steady-state populations}) exhibits familiar two-level system behaviour when considering the excited fraction
\begin{equation}
\frac{n_{\Uparrow}}{n_{\uparrow}} = \frac{1}{T + \left(\frac{\alpha}{\Omega}\right)^2 N_1/D}.
 \label{eq:U over u}
\end{equation}
The term $T$ (Eq. \eqref{eq:auxiliary functions}) is identical to the denominator of a Lorentzian two-level system line shape with peak at $\delta = 0$, and the additional term $\propto (A/\Omega)^2$ gives rise to hyperfine-induced broadening. The function $D$ can be interpreted as the denominator of a Lorentzian line shape centered where the frequency of the driving field matches the optical transition energy minus the Zeeman energy ($\delta = \omega$). The detunings $\delta = 0$ and $\delta = \omega$ are therefore of special importance, and we refer to them as optical and indirect resonances, respectively. If the laser frequency is far detuned from both of these resonances, then the term $TD \propto \Delta^4$ dominates, and the system tends towards the state $n_{\downarrow} = n_{\uparrow} = 1/2$ that is also obtained in the absence of optical driving.

We begin our investigation of the nuclear spin behaviour in the regime near optical resonance $\delta \lessapprox \Gamma$. Considering the ratio of ground state populations under the assumption of large Zeeman splitting appropriate for experiments in large ($\gtrapprox 1$T) magnetic fields, i.e. $\omega_e \gg A,\ \Omega,\ \eta_{\uparrow},\ \eta_{\downarrow}$, we arrive at
\begin{equation}
\frac{n_{\downarrow}}{n_{\uparrow}} \approx 1 + \frac{\Gamma^{\prime}}{\eta_t}\frac{1}{T}.
 \label{eq:ground state ration approx}
\end{equation}
The ratio of ground state populations is therefore described by almost the same Lorentzian function that gives the excited fraction in the optically driven two-level system obtained by switching off the hyperfine interaction. This Lorentzian feature is weighted by the ratio of spontaneous emission and combined dephasing $\Gamma^{\prime} / \eta_t = (\Gamma + \gamma)/(2\eta + \kappa_{\uparrow} + \kappa_{\downarrow})$, such that for slow dephasing population is strongly driven towards the state $\ket{\downarrow,m}$ which does not directly couple to the laser. For fast dephasing, on the other hand, the ground states are almost equally populated.

The nuclear spin steady state is obtained by calculating the spin-flip rates $r_{\uparrow / \downarrow}^{m_F}$ given by the solution of the one-block problem. In the regime of fast combined dephasing, the ground states can be assumed almost equally populated, and the trion state population follows an approximately Lorentzian line shape centered at the optical resonance. Hence given that the co-tunneling rates obey $|\kappa_{\uparrow} - \kappa_{\downarrow}| < \gamma$, the downward spin-flip rate $r_{\downarrow}(m_F) = \kappa_{\uparrow} n_{\uparrow}(m_F) + \gamma n_{\Uparrow}(m_F)$ dominates the upward spin-flip rate $r_{\uparrow}(m_F) = \kappa_{\downarrow}n_{\downarrow}(m_F)$ around the value of $m_F$ corresponding to the optical resonance due to the excited state contribution. Due to the negative g-factors in self-assembled quantum dots, the $\ket{\uparrow}$ electron spin state is usually lower in energy than the $\ket{\downarrow}$ spin state \cite{warburton_single_2013}, such that in the thermal state $n_{\uparrow} > n_{\downarrow}$, and we therefore take $\kappa_{\downarrow} > \kappa_{\uparrow}$. Hence for spin polarizations $m_F$ far from the value corresponding to the optical resonance where $n_{\Uparrow}(m_F) \approx 0$, and given that $n_{\uparrow}(m_F) \approx n_{\downarrow}(m_F)$, we find that the spin-flip rate upward dominates, i.e. $r_{\uparrow} > r_{\downarrow}$.

\begin{figure}
  \includegraphics[width=\linewidth]{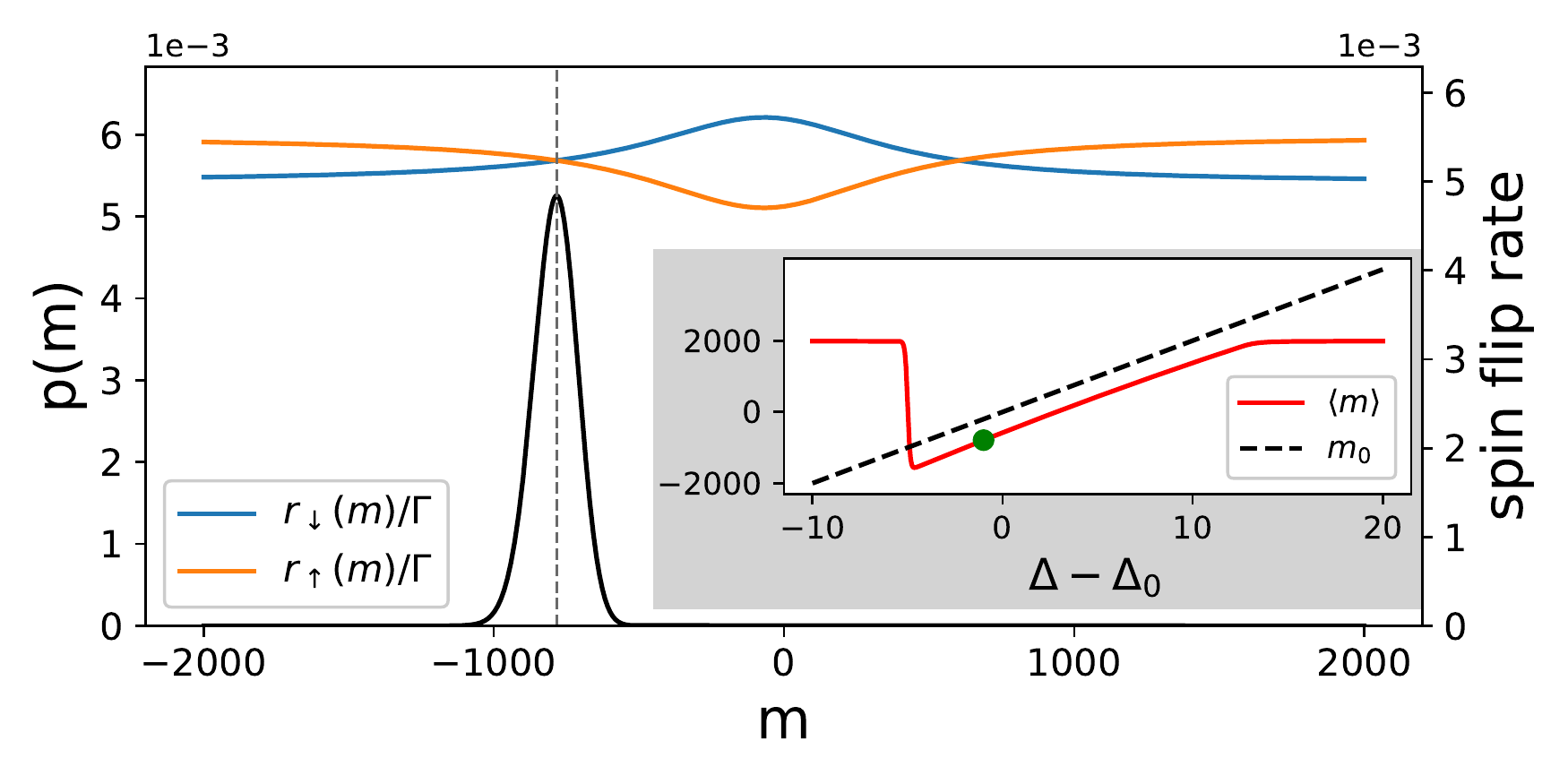}
 \caption{Spin-flip rates $r_{\downarrow} = \kappa_{\uparrow} n_{\uparrow} + \gamma n_{\Uparrow}$ and $r_{\uparrow} = \kappa_{\downarrow} n_{\downarrow}$ and resulting nuclear spin polarization probability distribution $p(m)$ (NSPPD, black line). Nuclear spin polarizations $m$ at which the rates match are the potential positions of a peak in the NSPPD.  Inset: The peak in the NSPPD remains at a fixed distance from the optical resonance polarization $m_0 = \frac{1}{A}(2\Delta - \omega_e)$ over a certain interval of detunings $\Delta$, which corresponds to the width of the flat-top line shape shown in Fig. \ref{fig:line shape}. The data point highlighted in green corresponds to the NSPPD shown. Parameters as in Fig. \ref{fig:line shape} with $\Delta = -74$ GHz for the outer figure.}
\label{fig:line dragging}
\end{figure}

There are two values $m_p$, $m_p^{\prime}$ of $m_F$ on the wings of the Lorentzian function given in Eq. \eqref{eq:U over u} at which the spin-flip rates are equal (see Fig. \ref{fig:line dragging}). The crossing point $m_p$ at the lower value of $m_F$ is a stable point, as higher (lower) $m_F$ values favour the spin-flip rate downward (upward), and the nuclear spin polarization is therefore driven towards this crossing point from both sides. The trapping condition $(r_{\uparrow} - r_{\downarrow})(m_p - m) > 1$ holds true for $m \in [-J,m_p^{\prime}]$, while polarization is pushed towards $m = +J$ for $m > m_p^{\prime}$. Clearly the resonance polarization $m_0 = \frac{1}{A}(2 \Delta - \omega_e)$ (peak of $r_{\downarrow}$ in Fig. \ref{fig:line dragging}) moves with the detuning $\Delta$ and with it the two crossing points $m_p$, $m_p^{\prime}$. In an upsweep of the laser frequency, a jump of the average nuclear spin polarization $\braket{m}$ to the near-resonant value $m_p$ occurs at a critical detuning $\Delta$ at which the trapping region captures a sufficient fraction of the nuclear spin range $m \in [-J,J]$. As shown in the inset of Fig. \ref{fig:line dragging} the average nuclear spin polarization then follows the resonance polarization $m_0$ up to its maximum polarization, and the NSPPD remains sharply peaked at a fixed and small detuning. This behaviour can be interpreted as the line dragging effect seen in experiments \cite{latta_confluence_2009, hogele_dynamic_2012}. Clearly the values of the co-tunneling rates $\kappa_{\uparrow / \downarrow}$ are swapped when the direction of the magnetic field is reversed, explaining why line anti-dragging rather than line dragging is observed when the lower-energy (`red') transition is driven rather than the higher-energy (`blue') one. In this case, $r_\downarrow>r_\uparrow$ for all $m$, and thus the NSPPD becomes peaked at $m=-J$.

The line-dragging behaviour described above can be seen as a consequence of two competing physical processes. On the one hand optical excitation from the ground state $\ket{\uparrow , m}$ to the trion state $\ket{\Uparrow, m}$ leads to diagonal decay to the state $\ket{\downarrow,m}$. On its own this process shelves electron spin population into the electron spin state that is not optically addressed, as observed experimentally in Ref. \onlinecite{dreiser_optical_2008}. In the presence of dephasing, however, the shelved state $\ket{\downarrow,m}$ can flip-flop with the nuclear spin system to the state $\ket{\uparrow,m-1}$, as the dephasing environment provides or absorbs the Zeeman energy required for the flip flop. This cycle of optical excitation - diagonal decay - environment-assisted flip flop decreases the nuclear spin projection number $m$ by one in each iteration. Following Ref. \onlinecite{latta_confluence_2009} we refer to this process as the Overhauser effect (see the left panel of Fig. \ref{fig:OHrevOH}).

On the other hand, co-tunneling pushes the electron spin state towards a slightly polarized thermal state at cryogenic temperatures. We find $\braket{S_z}_{\mathrm{thermal}} = 0.065 $ at 4 K and 3 T with a g-factor of $g_e = 30\ \mathrm{\mu e V / T}$ \cite{warburton_single_2013}. The hyperfine interaction leads to equilibration between electron and nuclear spin polarization given a mechanism to overcome the energy conservation barrier in an external magnetic field \cite{meier_optical_2012}. In the absence of optical driving, this mechanism is provided by dephasing in our model. A state $\ket{\uparrow , m}$ decays to the mixed state $1/2(\ket{\uparrow, m}\bra{\uparrow , m} + \ket{\downarrow, m+1}\bra{\downarrow , m+1})$ with a decay rate of $ \gamma_{eq} = 2 \eta_t \alpha^2 / \omega^2$ for $\omega \gg A, \eta_t$, where $\alpha$ is the effective flip-flop strength, $\omega$ the total electron spin splitting, and $\eta_t = \eta_{\uparrow} + \eta_{\downarrow}$ the total combined dephasing rate. This effect could in principle lead to nuclear spin polarization even in the absence of optical driving (as seen for detunings $\Delta < \Delta_c$ in the inset of Fig. \ref{fig:line dragging}, where $\Delta_c$ marks the position of the jump), but since the equilibration takes place at a comparatively slow rate of $\gamma_{eq}$ in the kHz regime for experiments in magetic fields of $\sim 1$ T, the effect is superseded by nuclear spin diffusion (diffusion rate $T_d \approx 2.5$ ms measured in Ref. \onlinecite{braun_bistability_2006}). Off-resonant optical driving leads to an additional dephasing mechanism and thereby accelerates the spin equilibration process, such that the thermal electron spin polarization is transferred to the nuclear spin system. In a system with a ground state $\ket{\uparrow}$ significantly lower in energy than $\ket{\downarrow}$, the electron tends towards positive $\braket{S}_{\mathrm{thermal}}$, such that the nuclear spin system upon equilibration also tends towards a positive value of polarization, which is in the opposite direction to the Overhauser effect. Since the Overhauser effect relies on optical excitation to the trion state $\ket{\Uparrow}$, it dominates only close to resonance, while thermal equilibration of thermal electron spin polarization takes place further off resonance. Hence the most likely nuclear spin polarization is found on the shoulder of the Lorentzian line shape describing the trion population. If the transition $\ket{\downarrow} \leftrightarrow \ket{\Downarrow}$ instead of $\ket{\uparrow} \leftrightarrow \ket{\Uparrow}$ of is optically addressed (using $\sigma^-$ rather than $\sigma^+$ polarized light), however, the Overhauser effect polarizes nuclear spins in the same direction as the thermal polarization, such that the most likely nuclear spin polarization is far off-resonant. This is consistent with the anti-dragging effect observed in Ref. \onlinecite{hogele_dynamic_2012}. As noted above, reversing the magnetic field switches between line dragging and anti-dragging regardless of which optical transition is driven, since reversing the field swaps the transition frequencies. Our model predicts that line dragging always occurs when the blue transition (whether this is $\ket{\uparrow}\leftrightarrow\ket{\Uparrow}$ or $\ket{\downarrow}\leftrightarrow\ket{\Downarrow}$) is driven, while anti-dragging occurs when the red transition is addressed, which agrees with experimental observations \cite{latta_confluence_2009, hogele_dynamic_2012}.

\section{Zeeman suppression}

We now show that in the weak-dephasing regime the electron-nuclear spin system exhibits behaviour that is very different from the line dragging phenomenon just described and that has thus far been missed by other theoretical treatments. Slow combined dephasing $\kappa_{\uparrow / \downarrow} + \eta \ll \Gamma$ effectively turns off the Overhauser effect, as environment-assisted equilibration of electron and nuclear spin polarization cannot take place any more. Nonetheless we find strong optically induced nuclear spin polarization effects in this low-dephasing regime due to second-order effects that combine a spin flip flop with an optical process. These processes are illustrated in the middle and right panels of Fig. 4. Before we describe them in detail, we first point out some general trends that follow from the ground state ratio obtained from Eq. \eqref{eq: steady-state populations}. For slow dephasing rates $\kappa_{\uparrow / \downarrow}, \eta \ll \Gamma$ this ratio approaches
\begin{equation}
\frac{n_{\downarrow}}{n_{\uparrow}} = 1 - \frac{\Omega^2 + \omega (2 \delta - \omega)}{\Omega^2 + (\Gamma^{\prime})^2 / 4 + \delta^2 + \alpha^2},
 \label{eq:limit zero dephasing}
\end{equation}
which holds in the limit of zero combined dephasing $\kappa_{\uparrow / \downarrow} + \eta = 0$. At the optical resonance ($\delta = 0$) the system can be seen to tend towards the state $\ket{\downarrow,m}$ assuming $\omega \gg \Omega$. At the indirect resonance $\omega = \delta$, on the contrary, the state $\ket{\uparrow,m-1}$ is favored. Hence the electron spin is pushed towards the state that is off resonance with the driving field in either case as shown in Fig. \ref{fig:2dPlot}, no matter if this state is optically addressed ($\ket{\uparrow,m-1}$) or not ($\ket{\downarrow, m}$). We now describe the physical origin of these trends and show that it leads to a new type of DNP in this system.

\begin{figure}
\includegraphics[width=\linewidth]{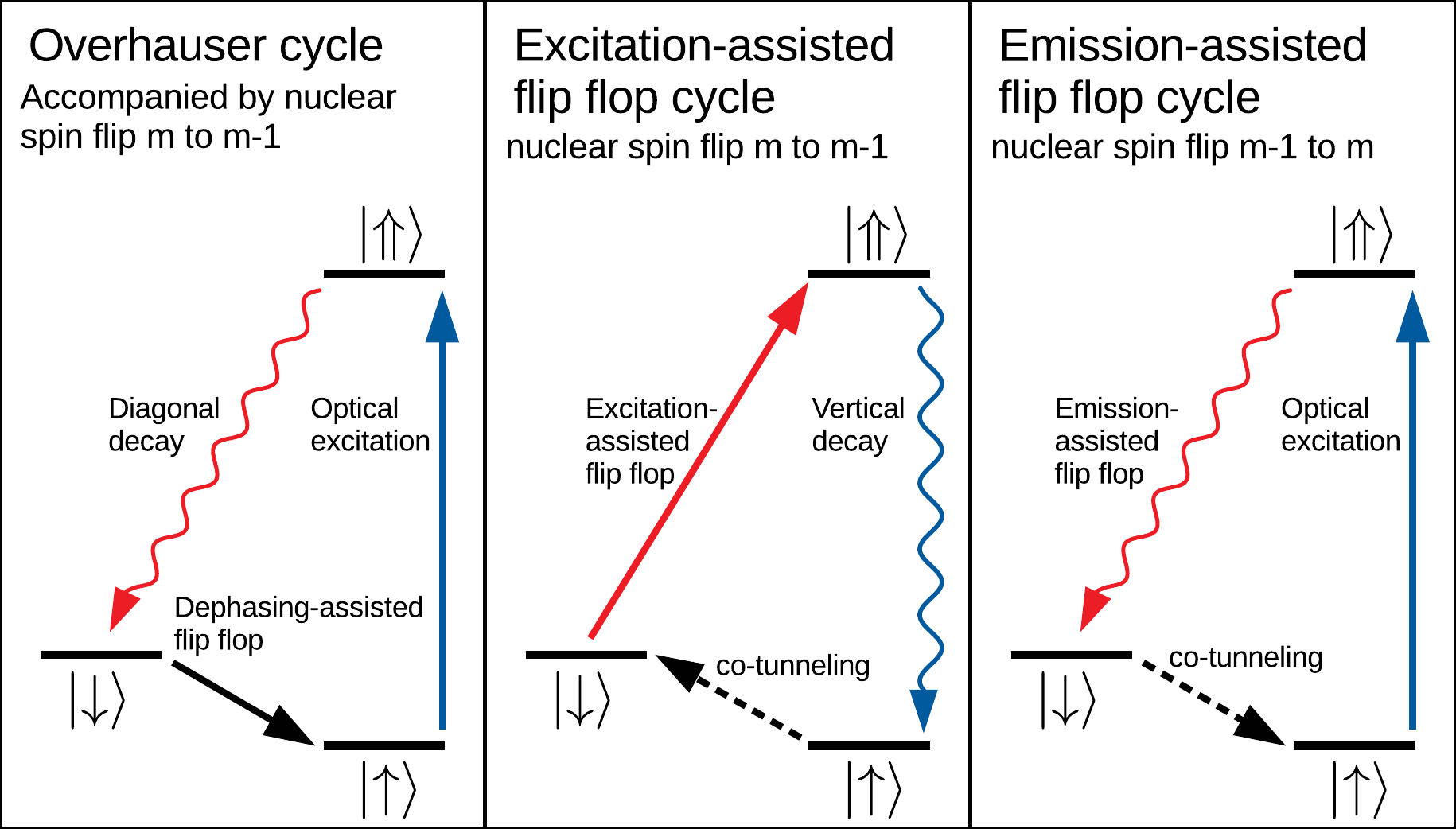}
\caption{Processes giving rise to optically induced nuclear polarization. The Overhauser cycle decreases nuclear spin polarization for a driven $\ket{\uparrow} \leftrightarrow \ket{\Uparrow}$ transition and requires diagonal decay as well as strong dephasing to overcome the energy conservation barrier of the electron-nuclear spin flip flop. The excitation- and emission-assisted flip flop cycles, on the contrary, take place even in the absence of diagonal decay and dominate the system dynamics in the weak dephasing regime, as the energy cost of the flip-flop is provided by an absorbed or emitted photon. The emission-assisted flip flop cycle is referred to as the reverse Overhauser effect in Ref. \onlinecite{latta_confluence_2009}. }
 \label{fig:OHrevOH}
\end{figure}

\begin{figure}
\begin{minipage}{0.9\linewidth}
\includegraphics[width=\linewidth]{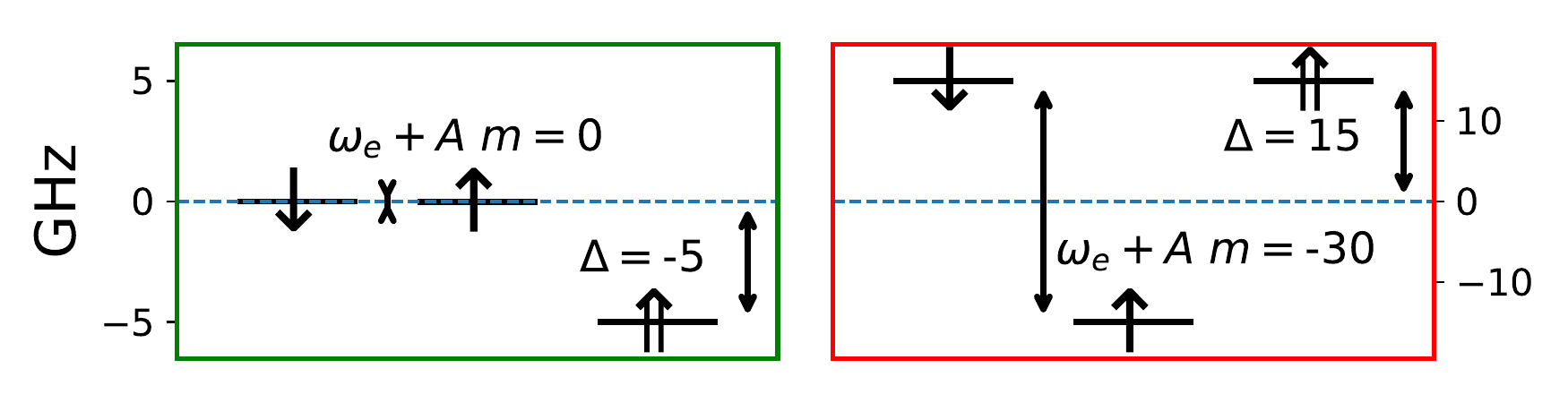}
\end{minipage}
\begin{minipage}{1.0\linewidth}
\includegraphics[width=\linewidth]{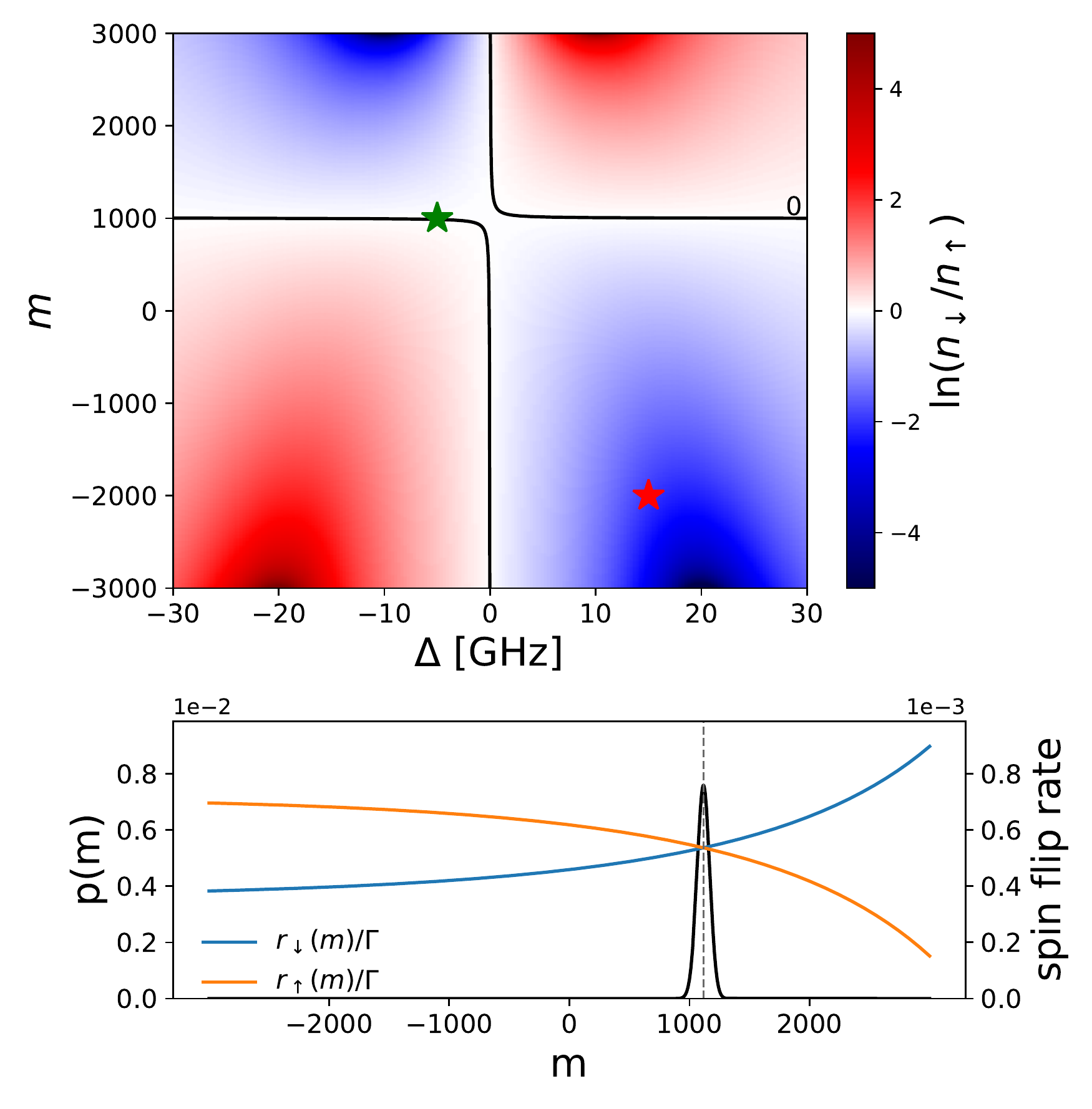}
\end{minipage}
\caption{
Top: Level diagrams in which the trion state $\ket{\Uparrow}$ is shown at the rotating frame energy $\Delta$ obtained by subtracting driving frequency from trion energy. The $m$ and $\Delta$ values for each diagram are indicated with a star of matching color in the centre plot.
Centre: The interplay of excitation- and emission-assisted flip flops can give rise to very unequal ground state populations of the optically driven three-level system, depending on nuclear spin polarization $m$ and detuning parameter $\Delta$. 
Bottom: Spin-flip rates $r_{\uparrow / \downarrow}$ (units of $\Gamma$) and resulting NSPPD vs. nuclear spin projection number $m$ for $\Delta = -5$ GHz, therefore corresponding to a vertical section through the point highlighted in green in the centre plot. The nuclear spins tend to polarize in such a way as to cancel the effect of the external magnetic field (Zeeman suppression). Parameters in GHz: $\Gamma = 1$, $\gamma = 0.01$, $A=0.01$, $\Omega = 1$, $\omega_e = -10$, $J=3000$, $\kappa_{\uparrow} = 0.9\ \kappa_{\downarrow} = \eta = 10^{-4}$.}
\label{fig:2dPlot} 
\end{figure}

Depletion of the $\ket{\uparrow,m-1}$ state near the optical resonance can be attributed to emission-assisted flip flops (right panel of Fig. \ref{fig:OHrevOH}). For small detunings $\delta$ the state $\ket{\uparrow,m-1}$ is likely to be excited to the trion state $\ket{\Uparrow,m-1}$ (Eq. \eqref{eq:U over u}).
In the presence of hyperfine coupling between the ground states $\ket{\downarrow,m}$, $\ket{\uparrow, m-1}$, the trion does not decay to $\ket{\uparrow,m-1}$ as the selection rules might suggest. Instead the decay results in a superposition of the energy eigenstates $\ket{\tilde{\uparrow}} = c_1 \ket{\uparrow , m-1} + c_2 \ket{\downarrow,m}$ and $\ket{\tilde{\downarrow}} = c_2 \ket{\uparrow , m-1} - c_1 \ket{\downarrow,m}$ with coefficients $c_{1,2}$ obtained by diagonalizing the sum of the hyperfine and Zeeman Hamiltonians. The state after photon emission is given by

\begin{equation}
\ket{\Uparrow,m-1} \xrightarrow[]{t\gg 1/\Gamma} c_1 \ket{\tilde{\uparrow}}\otimes \ket{\gamma_1} + c_2 \ket{\tilde{\downarrow}}\otimes \ket{\gamma_2},
 \label{eq:wavepacket result}
\end{equation}
where $\ket{\gamma_{1/2}}$ refer to single-photon wavepackets with a Lorentzian spectrum centered around the difference between trion energy and the energy of $\ket{\tilde{\uparrow} / \tilde{\downarrow}}$, respectively. Tracing over the photon and transforming to the original basis shows that a spontaneous-emission-assisted flip flop takes place with probability
\begin{equation}
p_{\Uparrow \rightarrow \downarrow} = \frac{1}{2}\frac{\left(\frac{\xi}{\alpha}\right)^2}{1 + \left(\frac{\xi}{2\alpha}\right)^2},
 \label{eq: emission assisted flipflop probability}
\end{equation}
where $\alpha = \frac{A}{2}\sqrt{J(J+1) - m(m-1)}$ as above and $\xi \equiv \sqrt{4\mathcal{A}^2 - (\omega_e + Am)^2} - (\omega_e + Am)$. As there are no energy conserving transitions from state $\ket{\downarrow, m}$, population accumulates in this state if the $\ket{\uparrow , m-1} \leftrightarrow \ket{\Uparrow, m-1}$ transition is driven resonantly in accordance with Eq. \eqref{eq:limit zero dephasing}. 

Depletion of the $\ket{\downarrow, m}$ state near the indirect resonance $\delta = \omega$ is a consequence of excitation-assisted flip flops (middle panel of Fig. \ref{fig:OHrevOH}). Formally this process can be seen by performing a Schrieffer-Wolff transformation to an effective Hamiltonian $\tilde{H} = \exp (S) H \exp (-S)$. Following Ref. \onlinecite{latta_confluence_2009}, we use $S = \frac{A}{2 \omega_e} \left( S^+ I^- - S^-I^+ \right)$, where as before $I^{\pm} = \sum_k I_k^{\pm}$, to arrive at
\begin{equation}\begin{split}
\tilde{H} = & \left( \omega_e + A I^z \right)S_z + \Omega \left( \ket{\uparrow}\bra{\Uparrow} + \ket{\Uparrow}\bra{\uparrow} \right) + \Delta \ket{\Uparrow}\bra{\Uparrow} \\
& - \frac{A \Omega}{2 \omega_e} \left( \ket{\downarrow} \bra{\Uparrow} I^+ + \ket{\Uparrow}\bra{\downarrow} I^- \right) + \mathcal{O}\left(\frac{A^2}{\omega_e}\right).
\end{split} \label{eq:SW effective Hamiltonian}
\end{equation}
The coupling term between states $\ket{\downarrow,m}$ and $\ket{\Uparrow, m-1}$ can be interpreted as the result of a second-order process involving a flip flop from state $\ket{\downarrow,m}$ to state $\ket{\uparrow,m-1}$ and photon emission into or absorption from the laser field connecting state $\ket{\uparrow,m-1}$ to $\ket{\Uparrow,m-1}$. We refer to this process as an excitation-assisted flip flop. It is resonant if the energy of a laser photon matches the energy of the $\ket{\uparrow, m-1} \leftrightarrow \ket{\Uparrow, m-1}$ transition minus the Zeeman energy cost of a flip flop, which corresponds to $\delta = \omega$ using the symbols of Eq. \eqref{eq:Hamiltonian aux symbols}. For driving near the indirect resonance $\delta = \omega$, the state $\ket{\downarrow,m}$ is therefore likely to undergo an excitation-assisted flip flop to trion state $\ket{\Uparrow,m-1}$, which then most probably decays to the ground state $\ket{\uparrow,m-1}$. Due to the large detuning of the optical transition, this ground state cannot be excited back to the trion state nor does it flip flop to $\ket{\downarrow,m}$ due to energy conservation. Hence the system tends towards $\ket{\uparrow,m-1}$ for $\delta \approx \omega$.

Optically induced nuclear spin polarization in the low-dephasing regime results from a competition of emission- and excitation-enabled flip flops. The emission-enabled process dominates and leads to $n_{\downarrow}(m) > n_{\uparrow}(m-1)$ if the driving frequency parametrized by $\Delta$ is closer to the $\ket{\uparrow,m-1} \leftrightarrow \ket{\Uparrow,m-1}$ transition energy than to that of $\ket{\downarrow,m} \leftrightarrow \ket{\Uparrow,m-1}$, as in this case direct excitation is more efficient than flip flop-enabled excitation. This condition is satisfied if the total electron splitting $A m + \omega_e$ and the detuning parameter $\Delta$ are both positive or both negative. The centre panel of Fig. \ref{fig:2dPlot} illustrates this effect for the ground state ratio $n_\downarrow/n_\uparrow$ obtained from Eq. \eqref{eq: steady-state populations} for a specific set of parameters in the slow-dephasing regime; $\Delta$ and $Am+\omega_e$ have the same sign in the upper right and lower left regions of the plot, and it is clear that $n_\downarrow$ dominates in these regions. In the level schemes shown in the top panel of the figure, these cases correspond to diagrams where both $\ket{\uparrow}$ and $\ket{\Uparrow}$ are below or above the zero line. On the contrary excitation-enabled flip flops dominate and give rise to $n_{\downarrow}(m) < n_{\uparrow}(m-1)$ if the system is closer to the indirect than to the direct resonance, which is satisfied when  $\Delta(\omega_e + Am) < 0$. The regions to the upper left and lower right of the zero contour in Fig. \ref{fig:2dPlot} therefore correspond to the regime of dominant excitation-enabled flip flops.

As the excited state population $n_{\Uparrow}(m-1)$ is vanishingly small in the low-dephasing regime, its contribution to the spin-flip rate $r_{\downarrow}^{m_F} = \kappa_{\uparrow} n_{\uparrow}(m-1) + \gamma n_{\Uparrow}(m-1)$ (Eq. \eqref{eq:rates i.t.o. dmat els}) can be neglected unless we have fast diagonal decay and slow co-tunneling, i.e. $\gamma / \kappa_{\uparrow / \downarrow} \gg 1$. The ratio of spin-flip rates $r_{\downarrow}^{m_F} \approx \kappa_{\uparrow} n_{\uparrow}$, $r_{\uparrow}^{m_F} = \kappa_{\downarrow} n_{\downarrow}$ for a particular laser frequency $\Delta$ can therefore be read off from the centre panel of Fig. \ref{fig:2dPlot} (assuming $\kappa_{\uparrow}/\kappa_{\downarrow} \approx 1$ for the moment). For $\Delta < 0$ the spin-flip rate $r_{\uparrow}$ dominates if $m < -\omega_e /A$ due to the prevalence of emission-assisted over excitation-assisted flip flops, while $r_{\downarrow}$ dominates if $m > -\omega_e /A$. Since for both $m>m_p$ and $m<m_p$ the rate that drives the polarization towards $m_{p} = - \omega_e/A$ dominates, the polarization probability accumulates at this value as is evident in the NSPPD shown in the bottom panel of Fig. \ref{fig:2dPlot}. In contrast, for $\Delta>0$, the rate $r_{\uparrow}$ dominates for $m > m_p$ and $r_{\downarrow}$ for $m < m_p$, such that there is no stable point, and the polarization probability is pushed away to either side of $m_p$. Similarly to the anti-dragging behavior discussed above, the nuclear spin polarization piles up towards maximum polarization at $m + \pm J$, which is why we refer to the behavior found in the $\Delta > 0$ case as anti-dragging as well. Unequal co-tunneling rates $\kappa_{\uparrow} \neq \kappa_{\downarrow}$ shift the position of $m_p$ but do not alter the qualitative behaviour. We therefore find that the nuclear spin system tends towards the value of polarization $m_p = - \omega_e /A$ that cancels the effect of the external magnetic field and leads to a zero total electron spin splitting for $\Delta < 0$ but avoids this value of polarization for $\Delta > 0$. This effect is one of the key findings of the present work, and we refer to it as Zeeman suppression.

The experiment required to demonstrate Zeeman suppression is comparatively simple and involves a circularly polarized laser driving a negatively charged QD in a magnetic field oriented along the optical axis. The laser frequency needs to be set such that it is closer to resonance with the blue electronic transition than to the forbidden one, i.e. $\Delta < 0$. The transition moves towards a decreased electron spin splitting, resulting in a temporarily increased photoluminescence (PL) if the laser frequency is lower than the original transition energy, which could be observed as a short spike in PL intensity. However the transition does not seek a position close to resonance as in the line-dragging regime seen for higher combined dephasing rates, but instead continues towards the configuration where both allowed electron transitions $\ket{\uparrow} \leftrightarrow \ket{\Uparrow}$ and $\ket{\downarrow} \leftrightarrow \ket{\Downarrow}$ have the same transition energy. If the system is driven for a sufficiently long time, the NSPPD will become concentrated around $m_p = - \omega_e / A$, which can be verified by measuring the transition energies of the two allowed electronic transitions. Executing these measurements without disturbing the nuclear state is an experimental challenge but can be achieved. In particular the technique of spin echo measurements with intermediate electron spin inversion pulses as presented in \cite{stockill_quantum_2016} seems suitable for this purpose. Such a measurement could even gain information on the narrowing of the polarization probability distribution, which is of great interest to QD-based photonic devices. The effect of an anti-aligned nuclear spin  steady state is expected to vanish when the laser frequency passes the $\Delta = 0$ level, at which point no narrowed distributions can be expected. Zeeman suppression should occur in experiments where there is little dephasing noise on top of the nuclear dephasing effects, which is achieved at low temperatures that suppress phonon effects. Moderate co-tunneling at rates $1/T_d < \kappa_{\uparrow / \downarrow} \ll \Gamma$, where $1/T_d$ is the nuclear spin diffusion rate, are required to complete the flip flop cycles shown in Fig. \ref{fig:OHrevOH}. Furthermore a moderate magnetic field $\omega_e \sim 10 \Gamma$ is conducive as it suppresses quadrupolar effects \cite{stockill_quantum_2016} and leads to a steady state of significant polarization around $m_p = - \omega_e /A$. A large magnetic field, however, suppresses the second-order processes involved and thereby prevents Zeeman suppression. We note the striking similarity between Zeeman suppression and the nuclear spin steady-state polarization counteracting the Zeeman splitting observed in the experiment presented in \cite{braun_bistability_2006}, however further theoretical work is needed to take the quasiresonant excitation laser used in the experiment into account.

Experimental observation of Zeeman suppression would be of great significance for the theoretical understanding of QDs and pave the way towards their use in photonic quantum technologies. It would demonstrate a dynamic nuclear polarization effect that arises out of the competition of two optically-assisted electron-nuclear spin transfer processes and that undergoes a sharp transition to a narrowly peaked polarization probability distribution at the critical laser frequency corresponding to $\Delta = 0$. This narrowly peaked distribution is centered around the polarization where the two transitions are degenerate, which is a crucial requirement for the generation of entangled photons for quantum computing following the protocols presented in Ref. \onlinecite{lindner_proposal_2009, economou_optically_2010, buterakos_deterministic_2017, russo_photonic_2018} and experimentally demonstrated in Ref. \onlinecite{schwartz_deterministic_2016-1}. Realization of such devices is currently hindered by the fast inhomogeneous dephasing due to the random value of nuclear spin polarization found in each run of an experiment. Zeeman suppression transforms this random distribution into a narrowly peaked one and therefore greatly extends the inhomogeneous dephasing time, potentially removing a major stumbling block of QD-based photonic quantum computing and communication.

\section{Discussion and conclusion}
We have presented an exact steady-state solution for the electron-nuclear spin system subject to continuous wave laser driving in the box model approximation. One of our goals in this work is to understand which DNP effects are captured by this minimal model, which includes only the contact hyperfine interaction, electron Zeeman interaction, and optical driving. We have seen that not only are line dragging and anti-dragging reproduced, but a new phenomenon that we refer to as Zeeman suppression is also predicted by this model. Here, we discuss how extensions to this model might modify the results we have obtained.

The approximation of uniform electron-nuclear coupling constants $A_k = \mathcal{A} \nu_0 |\Psi (\mathbf{r}_k)|^2$, where $\mathcal{A}$ is the hyperfine constant and $\nu_0$ the volume of the crystal unit cell, neglects the spatial variation of the electron wavefunction $\Psi (\mathbf{r}_k)$ across the nuclear positions $\mathbf{r}_k$. Inhomogeneity of the coupling constants $A_k$ can play a crucial role in the electron-spin dynamics as shown in Refs.  \onlinecite{khaetskii_electron_2002, khaetskii_electron_2003}, as the electron spin in a small magnetic field undergoes flip flop oscillations with the nuclei, and the amplitude and frequency of these oscillations is given by the coupling constants. Averaging over an inhomogeneous distribution $A_k$ therefore gives a decay of the electron spin polarization, while a sinusoidal oscillation is predicted for the box model limit $A_k \rightarrow \mathcal{A}/N$. This effect, however, can be neglected if the magnetic field is large enough to prevent first-order flip flops, which is satisfied for $B_{\mathrm{ext}} \gtrsim 100$ mT \cite{cywinski_pure_2009}. Moreover, the effect of the inhomogeneous hyperfine couplings in second-order flip-flop processes is negligible in the regime of strong optical driving given by $\Omega \sim \Gamma \sim 1$ GHz that is assumed in the DNP effects discussed above, since the electron spin steady state is reached much faster than the electron spin decoherence due to the difference in coupling constants $A_k$ \cite{barnes_nonperturbative_2012}. While the box model does provide a reliable model of electron spin behaviour in high magnetic fields or under strong optical driving, the question remains whether the inhomogeneity of the coupling constants qualitatively alters the nuclear spin response to the driven electron spin dynamics. As varying coupling constants $A_k$ break the block diagonal structure of the Hamiltonian \cite{petrov_coupled_2009}, the decoupling procedure leading to the steady-state solution of Eq. \eqref{eq:all-blocks solution} cannot be used to answer this question. However the mechanisms of environment-assisted angular momentum transfer giving rise to the DNP effects discussed here can be expected to play a role even in the case of inhomogeneous couplings, such that the box model can serve as an insightful limiting case.

The optically driven box model we consider does not include any mechanism of nuclear spin depolarization other than its interaction with the electron spin. The primary mechanism for nuclear spin depolarization is the inter-nuclear dipole-dipole interaction \cite{deng_nuclear_2005}, which drives nuclear spin diffusion out of the quantum dot. This interaction is weak compared to the second-order hyperfine processes we consider. In GaAs for example, the nuclear dipolar interaction between nearest neighbors is on the order of 0.1 neV, whereas the second-order drive-assisted flip-flop process shown in Eq. \eqref{eq:SW effective Hamiltonian} is on the order of A$\Omega/\omega_e \sim 1$  neV, which is an order of magnitude larger. Thus, the dipolar-induced nuclear spin diffusion is much slower than the hyperfine processes we consider and should play a negligible role in the initial build-up of DNP, which is the main focus of our work. The nuclear dipolar spin diffusion process will, however, ultimately limit the total amount of DNP that is created in the case of anti-dragging, and it would be interesting to incorporate this mechanism in future work to obtain more quantitatively accurate results for this particular case. We do not expect spin diffusion to have a significant effect on line-dragging or Zeeman suppression because these effects entail only moderate levels of DNP. Moreover, hyperfine processes become more dominant over dipolar interactions at low magnetic fields, so the relative importance of dipolar interactions should be even less in the case of Zeeman suppression.

Another potentially important mechanism that is not included in our model is the quadrupolar interacation, which has been shown to play an important role in experiments on self-assembled \cite{stockill_quantum_2016, gangloff_quantum_2018} and gate-defined \cite{botzem_quadrupolar_2016} QDs. The quadrupolar interaction leads to an additional electron spin dephasing mechanism and has been used in a model of line dragging \cite{hogele_dynamic_2012}. An interesting open question is therefore whether the linedragging mechanism proposed here can be experimentally distinguished from DNP relying on the quadrupolar interaction. As the quadrupolar interaction arises from electric field gradients occuring in strained QDs, the DNP mechanism at play could be identified by measuring a change of DNP behaviour with the QD size, externally applied pressure, or other parameters affecting the strain profile of the QD. The strong dependence of the linedragging mechanism presented here on dephasing due to co-tunneling or phonon coupling provides an additional experimental characteristic that could be used to distinguish Overhauser-cycle based DNP from other DNP mechanisms. Another open question beyond the scope of this work is whether the quadrupolar interaction can be included in a steady-state treatment of the kind presented here, which could help to clarify the role of this interaction.

In addition to showing that line dragging and anti-dragging can be explained using only the contact hyperfine interaction, we uncovered the novel DNP effect of Zeeman suppression, which arises out of an interplay of two optically assisted flip-flop processes. One of these mechanisms (emission-assisted flip flops) has been previously considered \cite{latta_confluence_2009} as a process competing with the Overhauser effect. We find that the two cycles take effect at different dephasing strengths and that the excitation-assisted flip flop cycle needs to be taken into account as well. Experimental observation of Zeeman suppression would be of great significance as it could be a way to establish conditions favourable for the realization of a QD source of entangled photons. Our result can also be used to explore other DNP effects such as the spontaneous self-polarization effect predicted in Ref. \onlinecite{korenev_nuclear_2007} in QDs and other systems featuring a driven and coupled electron-nuclear spin system.

T.N. acknowledges support from the Frontiers in Quantum Technologies programme (EU grant agreement no. 317232). E.B. and S.E. acknowledge support from the NSF (grant no. 1741656).

\bibliography{Box_CW_paper}

\begin{thebibliography}{41}%
\makeatletter
\providecommand \@ifxundefined [1]{%
 \@ifx{#1\undefined}
}%
\providecommand \@ifnum [1]{%
 \ifnum #1\expandafter \@firstoftwo
 \else \expandafter \@secondoftwo
 \fi
}%
\providecommand \@ifx [1]{%
 \ifx #1\expandafter \@firstoftwo
 \else \expandafter \@secondoftwo
 \fi
}%
\providecommand \natexlab [1]{#1}%
\providecommand \enquote  [1]{``#1''}%
\providecommand \bibnamefont  [1]{#1}%
\providecommand \bibfnamefont [1]{#1}%
\providecommand \citenamefont [1]{#1}%
\providecommand \href@noop [0]{\@secondoftwo}%
\providecommand \href [0]{\begingroup \@sanitize@url \@href}%
\providecommand \@href[1]{\@@startlink{#1}\@@href}%
\providecommand \@@href[1]{\endgroup#1\@@endlink}%
\providecommand \@sanitize@url [0]{\catcode `\\12\catcode `\$12\catcode
  `\&12\catcode `\#12\catcode `\^12\catcode `\_12\catcode `\%12\relax}%
\providecommand \@@startlink[1]{}%
\providecommand \@@endlink[0]{}%
\providecommand \url  [0]{\begingroup\@sanitize@url \@url }%
\providecommand \@url [1]{\endgroup\@href {#1}{\urlprefix }}%
\providecommand \urlprefix  [0]{URL }%
\providecommand \Eprint [0]{\href }%
\providecommand \doibase [0]{http://dx.doi.org/}%
\providecommand \selectlanguage [0]{\@gobble}%
\providecommand \bibinfo  [0]{\@secondoftwo}%
\providecommand \bibfield  [0]{\@secondoftwo}%
\providecommand \translation [1]{[#1]}%
\providecommand \BibitemOpen [0]{}%
\providecommand \bibitemStop [0]{}%
\providecommand \bibitemNoStop [0]{.\EOS\space}%
\providecommand \EOS [0]{\spacefactor3000\relax}%
\providecommand \BibitemShut  [1]{\csname bibitem#1\endcsname}%
\let\auto@bib@innerbib\@empty
%</preamble>
\bibitem [{\citenamefont {Barnes}\ and\ \citenamefont
  {Economou}(2011)}]{barnes_electron-nuclear_2011}%
  \BibitemOpen
  \bibfield  {author} {\bibinfo {author} {\bibfnamefont {E.}~\bibnamefont
  {Barnes}}\ and\ \bibinfo {author} {\bibfnamefont {S.~E.}\ \bibnamefont
  {Economou}},\ }\href {\doibase 10.1103/PhysRevLett.107.047601} {\bibfield
  {journal} {\bibinfo  {journal} {Physical Review Letters}\ }\textbf {\bibinfo
  {volume} {107}},\ \bibinfo {pages} {047601} (\bibinfo {year}
  {2011})}\BibitemShut {NoStop}%
\bibitem [{\citenamefont {Economou}\ and\ \citenamefont
  {Barnes}(2014)}]{economou_theory_2014}%
  \BibitemOpen
  \bibfield  {author} {\bibinfo {author} {\bibfnamefont {S.~E.}\ \bibnamefont
  {Economou}}\ and\ \bibinfo {author} {\bibfnamefont {E.}~\bibnamefont
  {Barnes}},\ }\href {\doibase 10.1103/PhysRevB.89.165301} {\bibfield
  {journal} {\bibinfo  {journal} {Physical Review B}\ }\textbf {\bibinfo
  {volume} {89}},\ \bibinfo {pages} {165301} (\bibinfo {year}
  {2014})}\BibitemShut {NoStop}%
\bibitem [{\citenamefont {Beugeling}\ \emph {et~al.}(2016)\citenamefont
  {Beugeling}, \citenamefont {Uhrig},\ and\ \citenamefont
  {Anders}}]{beugeling_quantum_2016}%
  \BibitemOpen
  \bibfield  {author} {\bibinfo {author} {\bibfnamefont {W.}~\bibnamefont
  {Beugeling}}, \bibinfo {author} {\bibfnamefont {G.~S.}\ \bibnamefont
  {Uhrig}}, \ and\ \bibinfo {author} {\bibfnamefont {F.~B.}\ \bibnamefont
  {Anders}},\ }\href@noop {} {\bibfield  {journal} {\bibinfo  {journal} {arXiv
  preprint arXiv:1609.06528}\ } (\bibinfo {year} {2016})}\BibitemShut {NoStop}%
\bibitem [{\citenamefont {Beugeling}\ \emph {et~al.}(2017)\citenamefont
  {Beugeling}, \citenamefont {Uhrig},\ and\ \citenamefont
  {Anders}}]{beugeling_influence_2017}%
  \BibitemOpen
  \bibfield  {author} {\bibinfo {author} {\bibfnamefont {W.}~\bibnamefont
  {Beugeling}}, \bibinfo {author} {\bibfnamefont {G.~S.}\ \bibnamefont
  {Uhrig}}, \ and\ \bibinfo {author} {\bibfnamefont {F.~B.}\ \bibnamefont
  {Anders}},\ }\href {\doibase 10.1103/PhysRevB.96.115303} {\bibfield
  {journal} {\bibinfo  {journal} {Physical Review B}\ }\textbf {\bibinfo
  {volume} {96}},\ \bibinfo {pages} {115303} (\bibinfo {year}
  {2017})}\BibitemShut {NoStop}%
\bibitem [{\citenamefont {Greilich}\ \emph {et~al.}(2007)\citenamefont
  {Greilich}, \citenamefont {Silva}, \citenamefont {Moussa}, \citenamefont
  {Ryan}, \citenamefont {Laforest}, \citenamefont {Baugh}, \citenamefont
  {Cory},\ and\ \citenamefont {Laflamme}}]{greilich_nuclei-induced_2007}%
  \BibitemOpen
  \bibfield  {author} {\bibinfo {author} {\bibfnamefont {J.}~\bibnamefont
  {Greilich}}, \bibinfo {author} {\bibfnamefont {M.}~\bibnamefont {Silva}},
  \bibinfo {author} {\bibfnamefont {O.}~\bibnamefont {Moussa}}, \bibinfo
  {author} {\bibfnamefont {C.}~\bibnamefont {Ryan}}, \bibinfo {author}
  {\bibfnamefont {M.}~\bibnamefont {Laforest}}, \bibinfo {author}
  {\bibfnamefont {J.}~\bibnamefont {Baugh}}, \bibinfo {author} {\bibfnamefont
  {D.~G.}\ \bibnamefont {Cory}}, \ and\ \bibinfo {author} {\bibfnamefont
  {R.}~\bibnamefont {Laflamme}},\ }\href {\doibase 10.1126/science.1145699}
  {\bibfield  {journal} {\bibinfo  {journal} {Science}\ }\textbf {\bibinfo
  {volume} {317}},\ \bibinfo {pages} {1893} (\bibinfo {year}
  {2007})}\BibitemShut {NoStop}%
\bibitem [{\citenamefont {Carter}\ \emph {et~al.}(2009)\citenamefont {Carter},
  \citenamefont {Shabaev}, \citenamefont {Economou}, \citenamefont {Kennedy},
  \citenamefont {Bracker},\ and\ \citenamefont
  {Reinecke}}]{carter_directing_2009}%
  \BibitemOpen
  \bibfield  {author} {\bibinfo {author} {\bibfnamefont {S.~G.}\ \bibnamefont
  {Carter}}, \bibinfo {author} {\bibfnamefont {A.}~\bibnamefont {Shabaev}},
  \bibinfo {author} {\bibfnamefont {S.~E.}\ \bibnamefont {Economou}}, \bibinfo
  {author} {\bibfnamefont {T.~A.}\ \bibnamefont {Kennedy}}, \bibinfo {author}
  {\bibfnamefont {A.~S.}\ \bibnamefont {Bracker}}, \ and\ \bibinfo {author}
  {\bibfnamefont {T.~L.}\ \bibnamefont {Reinecke}},\ }\href {\doibase
  10.1103/PhysRevLett.102.167403} {\bibfield  {journal} {\bibinfo  {journal}
  {Physical Review Letters}\ }\textbf {\bibinfo {volume} {102}},\ \bibinfo
  {pages} {167403} (\bibinfo {year} {2009})}\BibitemShut {NoStop}%
\bibitem [{\citenamefont {Latta}\ \emph {et~al.}(2009)\citenamefont {Latta},
  \citenamefont {H\"ogele}, \citenamefont {Zhao}, \citenamefont {Vamivakas},
  \citenamefont {Maletinsky}, \citenamefont {Kroner}, \citenamefont {Dreiser},
  \citenamefont {Carusotto}, \citenamefont {Badolato}, \citenamefont {Schuh},
  \citenamefont {Wegscheider}, \citenamefont {Atature},\ and\ \citenamefont
  {Imamoglu}}]{latta_confluence_2009}%
  \BibitemOpen
  \bibfield  {author} {\bibinfo {author} {\bibfnamefont {C.}~\bibnamefont
  {Latta}}, \bibinfo {author} {\bibfnamefont {A.}~\bibnamefont {H\"ogele}},
  \bibinfo {author} {\bibfnamefont {Y.}~\bibnamefont {Zhao}}, \bibinfo {author}
  {\bibfnamefont {A.~N.}\ \bibnamefont {Vamivakas}}, \bibinfo {author}
  {\bibfnamefont {P.}~\bibnamefont {Maletinsky}}, \bibinfo {author}
  {\bibfnamefont {M.}~\bibnamefont {Kroner}}, \bibinfo {author} {\bibfnamefont
  {J.}~\bibnamefont {Dreiser}}, \bibinfo {author} {\bibfnamefont
  {I.}~\bibnamefont {Carusotto}}, \bibinfo {author} {\bibfnamefont
  {A.}~\bibnamefont {Badolato}}, \bibinfo {author} {\bibfnamefont
  {D.}~\bibnamefont {Schuh}}, \bibinfo {author} {\bibfnamefont
  {W.}~\bibnamefont {Wegscheider}}, \bibinfo {author} {\bibfnamefont
  {M.}~\bibnamefont {Atature}}, \ and\ \bibinfo {author} {\bibfnamefont
  {A.}~\bibnamefont {Imamoglu}},\ }\href {\doibase 10.1038/nphys1363}
  {\bibfield  {journal} {\bibinfo  {journal} {Nature Physics}\ }\textbf
  {\bibinfo {volume} {5}},\ \bibinfo {pages} {758} (\bibinfo {year}
  {2009})}\BibitemShut {NoStop}%
\bibitem [{\citenamefont {H\"ogele}\ \emph {et~al.}(2012)\citenamefont
  {H\"ogele}, \citenamefont {Kroner}, \citenamefont {Latta}, \citenamefont
  {Claassen}, \citenamefont {Carusotto}, \citenamefont {Bulutay},\ and\
  \citenamefont {Imamoglu}}]{hogele_dynamic_2012}%
  \BibitemOpen
  \bibfield  {author} {\bibinfo {author} {\bibfnamefont {A.}~\bibnamefont
  {H\"ogele}}, \bibinfo {author} {\bibfnamefont {M.}~\bibnamefont {Kroner}},
  \bibinfo {author} {\bibfnamefont {C.}~\bibnamefont {Latta}}, \bibinfo
  {author} {\bibfnamefont {M.}~\bibnamefont {Claassen}}, \bibinfo {author}
  {\bibfnamefont {I.}~\bibnamefont {Carusotto}}, \bibinfo {author}
  {\bibfnamefont {C.}~\bibnamefont {Bulutay}}, \ and\ \bibinfo {author}
  {\bibfnamefont {A.}~\bibnamefont {Imamoglu}},\ }\href {\doibase
  10.1103/PhysRevLett.108.197403} {\bibfield  {journal} {\bibinfo  {journal}
  {Physical Review Letters}\ }\textbf {\bibinfo {volume} {108}},\ \bibinfo
  {pages} {197403} (\bibinfo {year} {2012})}\BibitemShut {NoStop}%
\bibitem [{\citenamefont {Xu}\ \emph {et~al.}(2009)\citenamefont {Xu},
  \citenamefont {Yao}, \citenamefont {Sun}, \citenamefont {Steel},
  \citenamefont {Bracker}, \citenamefont {Gammon},\ and\ \citenamefont
  {Sham}}]{xu_optically_2009}%
  \BibitemOpen
  \bibfield  {author} {\bibinfo {author} {\bibfnamefont {X.}~\bibnamefont
  {Xu}}, \bibinfo {author} {\bibfnamefont {W.}~\bibnamefont {Yao}}, \bibinfo
  {author} {\bibfnamefont {B.}~\bibnamefont {Sun}}, \bibinfo {author}
  {\bibfnamefont {D.~G.}\ \bibnamefont {Steel}}, \bibinfo {author}
  {\bibfnamefont {A.~S.}\ \bibnamefont {Bracker}}, \bibinfo {author}
  {\bibfnamefont {D.}~\bibnamefont {Gammon}}, \ and\ \bibinfo {author}
  {\bibfnamefont {L.~J.}\ \bibnamefont {Sham}},\ }\href {\doibase
  10.1038/nature08120} {\bibfield  {journal} {\bibinfo  {journal} {Nature}\
  }\textbf {\bibinfo {volume} {459}},\ \bibinfo {pages} {1105} (\bibinfo {year}
  {2009})}\BibitemShut {NoStop}%
\bibitem [{\citenamefont {Ladd}\ \emph {et~al.}(2010)\citenamefont {Ladd},
  \citenamefont {Press}, \citenamefont {De~Greve}, \citenamefont {McMahon},
  \citenamefont {Friess}, \citenamefont {Schneider}, \citenamefont {Kamp},
  \citenamefont {H\"ofling}, \citenamefont {Forchel},\ and\ \citenamefont
  {Yamamoto}}]{ladd_pulsed_2010}%
  \BibitemOpen
  \bibfield  {author} {\bibinfo {author} {\bibfnamefont {T.~D.}\ \bibnamefont
  {Ladd}}, \bibinfo {author} {\bibfnamefont {D.}~\bibnamefont {Press}},
  \bibinfo {author} {\bibfnamefont {K.}~\bibnamefont {De~Greve}}, \bibinfo
  {author} {\bibfnamefont {P.~L.}\ \bibnamefont {McMahon}}, \bibinfo {author}
  {\bibfnamefont {B.}~\bibnamefont {Friess}}, \bibinfo {author} {\bibfnamefont
  {C.}~\bibnamefont {Schneider}}, \bibinfo {author} {\bibfnamefont
  {M.}~\bibnamefont {Kamp}}, \bibinfo {author} {\bibfnamefont {S.}~\bibnamefont
  {H\"ofling}}, \bibinfo {author} {\bibfnamefont {A.}~\bibnamefont {Forchel}},
  \ and\ \bibinfo {author} {\bibfnamefont {Y.}~\bibnamefont {Yamamoto}},\
  }\href {\doibase 10.1103/PhysRevLett.105.107401} {\bibfield  {journal}
  {\bibinfo  {journal} {Physical Review Letters}\ }\textbf {\bibinfo {volume}
  {105}},\ \bibinfo {pages} {107401} (\bibinfo {year} {2010})}\BibitemShut
  {NoStop}%
\bibitem [{\citenamefont {Braun}\ \emph {et~al.}(2006)\citenamefont {Braun},
  \citenamefont {Urbaszek}, \citenamefont {Amand}, \citenamefont {Marie},
  \citenamefont {Krebs}, \citenamefont {Eble}, \citenamefont {Lemaitre},\ and\
  \citenamefont {Voisin}}]{braun_bistability_2006}%
  \BibitemOpen
  \bibfield  {author} {\bibinfo {author} {\bibfnamefont {P.-F.}\ \bibnamefont
  {Braun}}, \bibinfo {author} {\bibfnamefont {B.}~\bibnamefont {Urbaszek}},
  \bibinfo {author} {\bibfnamefont {T.}~\bibnamefont {Amand}}, \bibinfo
  {author} {\bibfnamefont {X.}~\bibnamefont {Marie}}, \bibinfo {author}
  {\bibfnamefont {O.}~\bibnamefont {Krebs}}, \bibinfo {author} {\bibfnamefont
  {B.}~\bibnamefont {Eble}}, \bibinfo {author} {\bibfnamefont {A.}~\bibnamefont
  {Lemaitre}}, \ and\ \bibinfo {author} {\bibfnamefont {P.}~\bibnamefont
  {Voisin}},\ }\href {\doibase 10.1103/PhysRevB.74.245306} {\bibfield
  {journal} {\bibinfo  {journal} {Physical Review B}\ }\textbf {\bibinfo
  {volume} {74}},\ \bibinfo {pages} {245306} (\bibinfo {year}
  {2006})}\BibitemShut {NoStop}%
\bibitem [{\citenamefont {Eble}\ \emph {et~al.}(2006)\citenamefont {Eble},
  \citenamefont {Krebs}, \citenamefont {Lema\^itre}, \citenamefont {Kowalik},
  \citenamefont {Kudelski}, \citenamefont {Voisin}, \citenamefont {Urbaszek},
  \citenamefont {Marie},\ and\ \citenamefont {Amand}}]{eble_dynamic_2006}%
  \BibitemOpen
  \bibfield  {author} {\bibinfo {author} {\bibfnamefont {B.}~\bibnamefont
  {Eble}}, \bibinfo {author} {\bibfnamefont {O.}~\bibnamefont {Krebs}},
  \bibinfo {author} {\bibfnamefont {A.}~\bibnamefont {Lema\^itre}}, \bibinfo
  {author} {\bibfnamefont {K.}~\bibnamefont {Kowalik}}, \bibinfo {author}
  {\bibfnamefont {A.}~\bibnamefont {Kudelski}}, \bibinfo {author}
  {\bibfnamefont {P.}~\bibnamefont {Voisin}}, \bibinfo {author} {\bibfnamefont
  {B.}~\bibnamefont {Urbaszek}}, \bibinfo {author} {\bibfnamefont
  {X.}~\bibnamefont {Marie}}, \ and\ \bibinfo {author} {\bibfnamefont
  {T.}~\bibnamefont {Amand}},\ }\href {\doibase 10.1103/PhysRevB.74.081306}
  {\bibfield  {journal} {\bibinfo  {journal} {Physical Review B}\ }\textbf
  {\bibinfo {volume} {74}},\ \bibinfo {pages} {081306} (\bibinfo {year}
  {2006})}\BibitemShut {NoStop}%
\bibitem [{\citenamefont {Krebs}\ \emph {et~al.}(2008)\citenamefont {Krebs},
  \citenamefont {Eble}, \citenamefont {Lema\^itre}, \citenamefont {Voisin},
  \citenamefont {Urbaszek}, \citenamefont {Amand},\ and\ \citenamefont
  {Marie}}]{krebs_hyperfine_2008}%
  \BibitemOpen
  \bibfield  {author} {\bibinfo {author} {\bibfnamefont {O.}~\bibnamefont
  {Krebs}}, \bibinfo {author} {\bibfnamefont {B.}~\bibnamefont {Eble}},
  \bibinfo {author} {\bibfnamefont {A.}~\bibnamefont {Lema\^itre}}, \bibinfo
  {author} {\bibfnamefont {P.}~\bibnamefont {Voisin}}, \bibinfo {author}
  {\bibfnamefont {B.}~\bibnamefont {Urbaszek}}, \bibinfo {author}
  {\bibfnamefont {T.}~\bibnamefont {Amand}}, \ and\ \bibinfo {author}
  {\bibfnamefont {X.}~\bibnamefont {Marie}},\ }\href {\doibase
  10.1016/j.crhy.2008.10.001} {\bibfield  {journal} {\bibinfo  {journal}
  {Comptes Rendus Physique}\ }\textbf {\bibinfo {volume} {9}},\ \bibinfo
  {pages} {874} (\bibinfo {year} {2008})}\BibitemShut {NoStop}%
\bibitem [{\citenamefont {Lai}\ \emph {et~al.}(2006)\citenamefont {Lai},
  \citenamefont {Maletinsky}, \citenamefont {Badolato},\ and\ \citenamefont
  {Imamoglu}}]{lai_knight-field-enabled_2006}%
  \BibitemOpen
  \bibfield  {author} {\bibinfo {author} {\bibfnamefont {C.~W.}\ \bibnamefont
  {Lai}}, \bibinfo {author} {\bibfnamefont {P.}~\bibnamefont {Maletinsky}},
  \bibinfo {author} {\bibfnamefont {A.}~\bibnamefont {Badolato}}, \ and\
  \bibinfo {author} {\bibfnamefont {A.}~\bibnamefont {Imamoglu}},\ }\href
  {\doibase 10.1103/PhysRevLett.96.167403} {\bibfield  {journal} {\bibinfo
  {journal} {Physical Review Letters}\ }\textbf {\bibinfo {volume} {96}},\
  \bibinfo {pages} {167403} (\bibinfo {year} {2006})}\BibitemShut {NoStop}%
\bibitem [{\citenamefont {Urbaszek}\ \emph {et~al.}(2013)\citenamefont
  {Urbaszek}, \citenamefont {Marie}, \citenamefont {Amand}, \citenamefont
  {Krebs}, \citenamefont {Voisin}, \citenamefont {Maletinsky}, \citenamefont
  {H\"ogele},\ and\ \citenamefont {Imamoglu}}]{urbaszek_nuclear_2013}%
  \BibitemOpen
  \bibfield  {author} {\bibinfo {author} {\bibfnamefont {B.}~\bibnamefont
  {Urbaszek}}, \bibinfo {author} {\bibfnamefont {X.}~\bibnamefont {Marie}},
  \bibinfo {author} {\bibfnamefont {T.}~\bibnamefont {Amand}}, \bibinfo
  {author} {\bibfnamefont {O.}~\bibnamefont {Krebs}}, \bibinfo {author}
  {\bibfnamefont {P.}~\bibnamefont {Voisin}}, \bibinfo {author} {\bibfnamefont
  {P.}~\bibnamefont {Maletinsky}}, \bibinfo {author} {\bibfnamefont
  {A.}~\bibnamefont {H\"ogele}}, \ and\ \bibinfo {author} {\bibfnamefont
  {A.}~\bibnamefont {Imamoglu}},\ }\href {\doibase 10.1103/RevModPhys.85.79}
  {\bibfield  {journal} {\bibinfo  {journal} {Reviews of Modern Physics}\
  }\textbf {\bibinfo {volume} {85}},\ \bibinfo {pages} {79} (\bibinfo {year}
  {2013})}\BibitemShut {NoStop}%
\bibitem [{\citenamefont {Wang}\ \emph {et~al.}(2018)\citenamefont {Wang},
  \citenamefont {Li}, \citenamefont {Jiang}, \citenamefont {He}, \citenamefont
  {Li}, \citenamefont {Ding}, \citenamefont {Chen}, \citenamefont {Qin},
  \citenamefont {Peng}, \citenamefont {Schneider}, \citenamefont {Kamp},
  \citenamefont {Zhang}, \citenamefont {Li}, \citenamefont {You}, \citenamefont
  {Wang}, \citenamefont {Dowling}, \citenamefont {H\"ofling}, \citenamefont
  {Lu},\ and\ \citenamefont {Pan}}]{wang_toward_2018}%
  \BibitemOpen
  \bibfield  {author} {\bibinfo {author} {\bibfnamefont {H.}~\bibnamefont
  {Wang}}, \bibinfo {author} {\bibfnamefont {W.}~\bibnamefont {Li}}, \bibinfo
  {author} {\bibfnamefont {X.}~\bibnamefont {Jiang}}, \bibinfo {author}
  {\bibfnamefont {Y.-M.}\ \bibnamefont {He}}, \bibinfo {author} {\bibfnamefont
  {Y.-H.}\ \bibnamefont {Li}}, \bibinfo {author} {\bibfnamefont
  {X.}~\bibnamefont {Ding}}, \bibinfo {author} {\bibfnamefont {M.-C.}\
  \bibnamefont {Chen}}, \bibinfo {author} {\bibfnamefont {J.}~\bibnamefont
  {Qin}}, \bibinfo {author} {\bibfnamefont {C.-Z.}\ \bibnamefont {Peng}},
  \bibinfo {author} {\bibfnamefont {C.}~\bibnamefont {Schneider}}, \bibinfo
  {author} {\bibfnamefont {M.}~\bibnamefont {Kamp}}, \bibinfo {author}
  {\bibfnamefont {W.-J.}\ \bibnamefont {Zhang}}, \bibinfo {author}
  {\bibfnamefont {H.}~\bibnamefont {Li}}, \bibinfo {author} {\bibfnamefont
  {L.-X.}\ \bibnamefont {You}}, \bibinfo {author} {\bibfnamefont
  {Z.}~\bibnamefont {Wang}}, \bibinfo {author} {\bibfnamefont {J.~P.}\
  \bibnamefont {Dowling}}, \bibinfo {author} {\bibfnamefont {S.}~\bibnamefont
  {H\"ofling}}, \bibinfo {author} {\bibfnamefont {C.-Y.}\ \bibnamefont {Lu}}, \
  and\ \bibinfo {author} {\bibfnamefont {J.-W.}\ \bibnamefont {Pan}},\ }\href
  {\doibase 10.1103/PhysRevLett.120.230502} {\bibfield  {journal} {\bibinfo
  {journal} {Physical Review Letters}\ }\textbf {\bibinfo {volume} {120}},\
  \bibinfo {pages} {230502} (\bibinfo {year} {2018})}\BibitemShut {NoStop}%
\bibitem [{\citenamefont {Hanschke}\ \emph {et~al.}(2018)\citenamefont
  {Hanschke}, \citenamefont {Fischer}, \citenamefont {Appel}, \citenamefont
  {Lukin}, \citenamefont {Wierzbowski}, \citenamefont {Sun}, \citenamefont
  {Trivedi}, \citenamefont {Vu{\v c}kovi\'c}, \citenamefont {Finley},\ and\
  \citenamefont {M\"uller}}]{hanschke_quantum_2018}%
  \BibitemOpen
  \bibfield  {author} {\bibinfo {author} {\bibfnamefont {L.}~\bibnamefont
  {Hanschke}}, \bibinfo {author} {\bibfnamefont {K.~A.}\ \bibnamefont
  {Fischer}}, \bibinfo {author} {\bibfnamefont {S.}~\bibnamefont {Appel}},
  \bibinfo {author} {\bibfnamefont {D.}~\bibnamefont {Lukin}}, \bibinfo
  {author} {\bibfnamefont {J.}~\bibnamefont {Wierzbowski}}, \bibinfo {author}
  {\bibfnamefont {S.}~\bibnamefont {Sun}}, \bibinfo {author} {\bibfnamefont
  {R.}~\bibnamefont {Trivedi}}, \bibinfo {author} {\bibfnamefont
  {J.}~\bibnamefont {Vu{\v c}kovi\'c}}, \bibinfo {author} {\bibfnamefont
  {J.~J.}\ \bibnamefont {Finley}}, \ and\ \bibinfo {author} {\bibfnamefont
  {K.}~\bibnamefont {M\"uller}},\ }\href {\doibase 10.1038/s41534-018-0092-0}
  {\bibfield  {journal} {\bibinfo  {journal} {npj Quantum Information}\
  }\textbf {\bibinfo {volume} {4}},\ \bibinfo {pages} {41534} (\bibinfo {year}
  {2018})}\BibitemShut {NoStop}%
\bibitem [{\citenamefont {Somaschi}\ \emph {et~al.}(2016)\citenamefont
  {Somaschi}, \citenamefont {Giesz}, \citenamefont {De~Santis}, \citenamefont
  {Loredo}, \citenamefont {Almeida}, \citenamefont {Hornecker}, \citenamefont
  {Portalupi}, \citenamefont {Grange}, \citenamefont {Ant\'on}, \citenamefont
  {Demory}, \citenamefont {G\'omez}, \citenamefont {Sagnes}, \citenamefont
  {{Lanzillotti-Kimura}}, \citenamefont {Lema\'itre}, \citenamefont {Auffeves},
  \citenamefont {White}, \citenamefont {Lanco},\ and\ \citenamefont
  {Senellart}}]{somaschi_near-optimal_2016}%
  \BibitemOpen
  \bibfield  {author} {\bibinfo {author} {\bibfnamefont {N.}~\bibnamefont
  {Somaschi}}, \bibinfo {author} {\bibfnamefont {V.}~\bibnamefont {Giesz}},
  \bibinfo {author} {\bibfnamefont {L.}~\bibnamefont {De~Santis}}, \bibinfo
  {author} {\bibfnamefont {J.~C.}\ \bibnamefont {Loredo}}, \bibinfo {author}
  {\bibfnamefont {M.~P.}\ \bibnamefont {Almeida}}, \bibinfo {author}
  {\bibfnamefont {G.}~\bibnamefont {Hornecker}}, \bibinfo {author}
  {\bibfnamefont {S.~L.}\ \bibnamefont {Portalupi}}, \bibinfo {author}
  {\bibfnamefont {T.}~\bibnamefont {Grange}}, \bibinfo {author} {\bibfnamefont
  {C.}~\bibnamefont {Ant\'on}}, \bibinfo {author} {\bibfnamefont
  {J.}~\bibnamefont {Demory}}, \bibinfo {author} {\bibfnamefont
  {C.}~\bibnamefont {G\'omez}}, \bibinfo {author} {\bibfnamefont
  {I.}~\bibnamefont {Sagnes}}, \bibinfo {author} {\bibfnamefont {N.~D.}\
  \bibnamefont {{Lanzillotti-Kimura}}}, \bibinfo {author} {\bibfnamefont
  {A.}~\bibnamefont {Lema\'itre}}, \bibinfo {author} {\bibfnamefont
  {A.}~\bibnamefont {Auffeves}}, \bibinfo {author} {\bibfnamefont {A.~G.}\
  \bibnamefont {White}}, \bibinfo {author} {\bibfnamefont {L.}~\bibnamefont
  {Lanco}}, \ and\ \bibinfo {author} {\bibfnamefont {P.}~\bibnamefont
  {Senellart}},\ }\href {\doibase 10.1038/nphoton.2016.23} {\bibfield
  {journal} {\bibinfo  {journal} {Nature Photonics}\ }\textbf {\bibinfo
  {volume} {10}},\ \bibinfo {pages} {340} (\bibinfo {year} {2016})}\BibitemShut
  {NoStop}%
\bibitem [{\citenamefont {Ding}\ \emph {et~al.}(2016)\citenamefont {Ding},
  \citenamefont {He}, \citenamefont {Duan}, \citenamefont {Gregersen},
  \citenamefont {Chen}, \citenamefont {Unsleber}, \citenamefont {Maier},
  \citenamefont {Schneider}, \citenamefont {Kamp}, \citenamefont {H\"ofling},
  \citenamefont {Lu},\ and\ \citenamefont {Pan}}]{ding_-demand_2016}%
  \BibitemOpen
  \bibfield  {author} {\bibinfo {author} {\bibfnamefont {X.}~\bibnamefont
  {Ding}}, \bibinfo {author} {\bibfnamefont {Y.}~\bibnamefont {He}}, \bibinfo
  {author} {\bibfnamefont {Z.-C.}\ \bibnamefont {Duan}}, \bibinfo {author}
  {\bibfnamefont {N.}~\bibnamefont {Gregersen}}, \bibinfo {author}
  {\bibfnamefont {M.-C.}\ \bibnamefont {Chen}}, \bibinfo {author}
  {\bibfnamefont {S.}~\bibnamefont {Unsleber}}, \bibinfo {author}
  {\bibfnamefont {S.}~\bibnamefont {Maier}}, \bibinfo {author} {\bibfnamefont
  {C.}~\bibnamefont {Schneider}}, \bibinfo {author} {\bibfnamefont
  {M.}~\bibnamefont {Kamp}}, \bibinfo {author} {\bibfnamefont {S.}~\bibnamefont
  {H\"ofling}}, \bibinfo {author} {\bibfnamefont {C.-Y.}\ \bibnamefont {Lu}}, \
  and\ \bibinfo {author} {\bibfnamefont {J.-W.}\ \bibnamefont {Pan}},\ }\href
  {\doibase 10.1103/PhysRevLett.116.020401} {\bibfield  {journal} {\bibinfo
  {journal} {Physical Review Letters}\ }\textbf {\bibinfo {volume} {116}},\
  \bibinfo {pages} {020401} (\bibinfo {year} {2016})}\BibitemShut {NoStop}%
\bibitem [{\citenamefont {Schwartz}\ \emph {et~al.}(2016)\citenamefont
  {Schwartz}, \citenamefont {Cogan}, \citenamefont {Schmidgall}, \citenamefont
  {Don}, \citenamefont {Gantz}, \citenamefont {Kenneth}, \citenamefont
  {Lindner},\ and\ \citenamefont {Gershoni}}]{schwartz_deterministic_2016-1}%
  \BibitemOpen
  \bibfield  {author} {\bibinfo {author} {\bibfnamefont {I.}~\bibnamefont
  {Schwartz}}, \bibinfo {author} {\bibfnamefont {D.}~\bibnamefont {Cogan}},
  \bibinfo {author} {\bibfnamefont {E.~R.}\ \bibnamefont {Schmidgall}},
  \bibinfo {author} {\bibfnamefont {Y.}~\bibnamefont {Don}}, \bibinfo {author}
  {\bibfnamefont {L.}~\bibnamefont {Gantz}}, \bibinfo {author} {\bibfnamefont
  {O.}~\bibnamefont {Kenneth}}, \bibinfo {author} {\bibfnamefont {N.~H.}\
  \bibnamefont {Lindner}}, \ and\ \bibinfo {author} {\bibfnamefont
  {D.}~\bibnamefont {Gershoni}},\ }\href {\doibase 10.1126/science.aah4758}
  {\bibfield  {journal} {\bibinfo  {journal} {Science}\ }\textbf {\bibinfo
  {volume} {354}},\ \bibinfo {pages} {434} (\bibinfo {year}
  {2016})}\BibitemShut {NoStop}%
\bibitem [{\citenamefont {Burkard}\ \emph {et~al.}(1999)\citenamefont
  {Burkard}, \citenamefont {Loss},\ and\ \citenamefont
  {DiVincenzo}}]{burkard_coupled_1999}%
  \BibitemOpen
  \bibfield  {author} {\bibinfo {author} {\bibfnamefont {G.}~\bibnamefont
  {Burkard}}, \bibinfo {author} {\bibfnamefont {D.}~\bibnamefont {Loss}}, \
  and\ \bibinfo {author} {\bibfnamefont {D.~P.}\ \bibnamefont {DiVincenzo}},\
  }\href@noop {} {\bibfield  {journal} {\bibinfo  {journal} {Physical Review
  B}\ }\textbf {\bibinfo {volume} {59}},\ \bibinfo {pages} {2070} (\bibinfo
  {year} {1999})}\BibitemShut {NoStop}%
\bibitem [{\citenamefont {Barnes}\ \emph {et~al.}(2011)\citenamefont {Barnes},
  \citenamefont {Cywi\'nski},\ and\ \citenamefont
  {Das~Sarma}}]{barnes_master_2011}%
  \BibitemOpen
  \bibfield  {author} {\bibinfo {author} {\bibfnamefont {E.}~\bibnamefont
  {Barnes}}, \bibinfo {author} {\bibfnamefont {L.}~\bibnamefont {Cywi\'nski}},
  \ and\ \bibinfo {author} {\bibfnamefont {S.}~\bibnamefont {Das~Sarma}},\
  }\href {\doibase 10.1103/PhysRevB.84.155315} {\bibfield  {journal} {\bibinfo
  {journal} {Physical Review B}\ }\textbf {\bibinfo {volume} {84}},\ \bibinfo
  {pages} {155315} (\bibinfo {year} {2011})}\BibitemShut {NoStop}%
\bibitem [{\citenamefont {Barnes}\ \emph {et~al.}(2012)\citenamefont {Barnes},
  \citenamefont {Cywi\'nski},\ and\ \citenamefont
  {Das~Sarma}}]{barnes_nonperturbative_2012}%
  \BibitemOpen
  \bibfield  {author} {\bibinfo {author} {\bibfnamefont {E.}~\bibnamefont
  {Barnes}}, \bibinfo {author} {\bibfnamefont {L.}~\bibnamefont {Cywi\'nski}},
  \ and\ \bibinfo {author} {\bibfnamefont {S.}~\bibnamefont {Das~Sarma}},\
  }\href {\doibase 10.1103/PhysRevLett.109.140403} {\bibfield  {journal}
  {\bibinfo  {journal} {Physical Review Letters}\ }\textbf {\bibinfo {volume}
  {109}},\ \bibinfo {pages} {140403} (\bibinfo {year} {2012})}\BibitemShut
  {NoStop}%
\bibitem [{\citenamefont {Lindner}\ and\ \citenamefont
  {Rudolph}(2009)}]{lindner_proposal_2009}%
  \BibitemOpen
  \bibfield  {author} {\bibinfo {author} {\bibfnamefont {N.~H.}\ \bibnamefont
  {Lindner}}\ and\ \bibinfo {author} {\bibfnamefont {T.}~\bibnamefont
  {Rudolph}},\ }\href {\doibase 10.1103/PhysRevLett.103.113602} {\bibfield
  {journal} {\bibinfo  {journal} {Physical Review Letters}\ }\textbf {\bibinfo
  {volume} {103}},\ \bibinfo {pages} {113602} (\bibinfo {year}
  {2009})}\BibitemShut {NoStop}%
\bibitem [{\citenamefont {Kozlov}(2007)}]{kozlov_exactly_2007}%
  \BibitemOpen
  \bibfield  {author} {\bibinfo {author} {\bibfnamefont {G.~G.}\ \bibnamefont
  {Kozlov}},\ }\href {\doibase 10.1134/S1063776107100159} {\bibfield  {journal}
  {\bibinfo  {journal} {Journal of Experimental and Theoretical Physics}\
  }\textbf {\bibinfo {volume} {105}},\ \bibinfo {pages} {803} (\bibinfo {year}
  {2007})}\BibitemShut {NoStop}%
\bibitem [{\citenamefont {Dreiser}\ \emph {et~al.}(2008)\citenamefont
  {Dreiser}, \citenamefont {Atat\"ure}, \citenamefont {Galland}, \citenamefont
  {M\"uller}, \citenamefont {Badolato},\ and\ \citenamefont
  {Imamoglu}}]{dreiser_optical_2008}%
  \BibitemOpen
  \bibfield  {author} {\bibinfo {author} {\bibfnamefont {J.}~\bibnamefont
  {Dreiser}}, \bibinfo {author} {\bibfnamefont {M.}~\bibnamefont {Atat\"ure}},
  \bibinfo {author} {\bibfnamefont {C.}~\bibnamefont {Galland}}, \bibinfo
  {author} {\bibfnamefont {T.}~\bibnamefont {M\"uller}}, \bibinfo {author}
  {\bibfnamefont {A.}~\bibnamefont {Badolato}}, \ and\ \bibinfo {author}
  {\bibfnamefont {A.}~\bibnamefont {Imamoglu}},\ }\href {\doibase
  10.1103/PhysRevB.77.075317} {\bibfield  {journal} {\bibinfo  {journal}
  {Physical Review B}\ }\textbf {\bibinfo {volume} {77}},\ \bibinfo {pages}
  {075317} (\bibinfo {year} {2008})}\BibitemShut {NoStop}%
\bibitem [{\citenamefont {Grosse}\ \emph {et~al.}(2008)\citenamefont {Grosse},
  \citenamefont {Muljarov},\ and\ \citenamefont
  {Zimmermann}}]{grosse_phonons_2008}%
  \BibitemOpen
  \bibfield  {author} {\bibinfo {author} {\bibfnamefont {F.}~\bibnamefont
  {Grosse}}, \bibinfo {author} {\bibfnamefont {E.}~\bibnamefont {Muljarov}}, \
  and\ \bibinfo {author} {\bibfnamefont {R.}~\bibnamefont {Zimmermann}},\ }in\
  \href@noop {} {\emph {\bibinfo {booktitle} {Semiconductor
  {{Nanostructures}}}}},\ \bibinfo {series and number} {{{NanoScience}} and
  {{Technology}}}\ (\bibinfo  {publisher} {{Springer}},\ \bibinfo {address}
  {Berlin, Heidelberg},\ \bibinfo {year} {2008})\BibitemShut {NoStop}%
\bibitem [{\citenamefont {Warburton}(2013)}]{warburton_single_2013}%
  \BibitemOpen
  \bibfield  {author} {\bibinfo {author} {\bibfnamefont {R.~J.}\ \bibnamefont
  {Warburton}},\ }\href {\doibase 10.1038/nmat3585} {\bibfield  {journal}
  {\bibinfo  {journal} {Nature Materials}\ }\textbf {\bibinfo {volume} {12}},\
  \bibinfo {pages} {483} (\bibinfo {year} {2013})}\BibitemShut {NoStop}%
\bibitem [{\citenamefont {Meier}\ and\ \citenamefont
  {Zakharchenya}(2012)}]{meier_optical_2012}%
  \BibitemOpen
  \bibfield  {author} {\bibinfo {author} {\bibfnamefont {F.}~\bibnamefont
  {Meier}}\ and\ \bibinfo {author} {\bibfnamefont {B.~P.}\ \bibnamefont
  {Zakharchenya}},\ }\href@noop {} {\emph {\bibinfo {title} {Optical
  {{Orientation}}}}},\ Modern {{Problems}} in {{Condensed Matter Sciences}}\
  (\bibinfo  {publisher} {{Elsevier Science}},\ \bibinfo {address}
  {Amsterdam},\ \bibinfo {year} {2012})\BibitemShut {NoStop}%
\bibitem [{\citenamefont {Stockill}\ \emph {et~al.}(2016)\citenamefont
  {Stockill}, \citenamefont {Le~Gall}, \citenamefont {Matthiesen},
  \citenamefont {Huthmacher}, \citenamefont {Clarke}, \citenamefont {Hugues},\
  and\ \citenamefont {Atat\"ure}}]{stockill_quantum_2016}%
  \BibitemOpen
  \bibfield  {author} {\bibinfo {author} {\bibfnamefont {R.}~\bibnamefont
  {Stockill}}, \bibinfo {author} {\bibfnamefont {C.}~\bibnamefont {Le~Gall}},
  \bibinfo {author} {\bibfnamefont {C.}~\bibnamefont {Matthiesen}}, \bibinfo
  {author} {\bibfnamefont {L.}~\bibnamefont {Huthmacher}}, \bibinfo {author}
  {\bibfnamefont {E.}~\bibnamefont {Clarke}}, \bibinfo {author} {\bibfnamefont
  {M.}~\bibnamefont {Hugues}}, \ and\ \bibinfo {author} {\bibfnamefont
  {M.}~\bibnamefont {Atat\"ure}},\ }\href {\doibase 10.1038/ncomms12745}
  {\bibfield  {journal} {\bibinfo  {journal} {Nature Communications}\ }\textbf
  {\bibinfo {volume} {7}},\ \bibinfo {pages} {12745} (\bibinfo {year}
  {2016})}\BibitemShut {NoStop}%
\bibitem [{\citenamefont {Economou}\ \emph {et~al.}(2010)\citenamefont
  {Economou}, \citenamefont {Lindner},\ and\ \citenamefont
  {Rudolph}}]{economou_optically_2010}%
  \BibitemOpen
  \bibfield  {author} {\bibinfo {author} {\bibfnamefont {S.~E.}\ \bibnamefont
  {Economou}}, \bibinfo {author} {\bibfnamefont {N.}~\bibnamefont {Lindner}}, \
  and\ \bibinfo {author} {\bibfnamefont {T.}~\bibnamefont {Rudolph}},\ }\href
  {\doibase 10.1103/PhysRevLett.105.093601} {\bibfield  {journal} {\bibinfo
  {journal} {Physical Review Letters}\ }\textbf {\bibinfo {volume} {105}},\
  \bibinfo {pages} {093601} (\bibinfo {year} {2010})}\BibitemShut {NoStop}%
\bibitem [{\citenamefont {Buterakos}\ \emph {et~al.}(2017)\citenamefont
  {Buterakos}, \citenamefont {Barnes},\ and\ \citenamefont
  {Economou}}]{buterakos_deterministic_2017}%
  \BibitemOpen
  \bibfield  {author} {\bibinfo {author} {\bibfnamefont {D.}~\bibnamefont
  {Buterakos}}, \bibinfo {author} {\bibfnamefont {E.}~\bibnamefont {Barnes}}, \
  and\ \bibinfo {author} {\bibfnamefont {S.~E.}\ \bibnamefont {Economou}},\
  }\href {\doibase 10.1103/PhysRevX.7.041023} {\bibfield  {journal} {\bibinfo
  {journal} {Physical Review X}\ }\textbf {\bibinfo {volume} {7}},\ \bibinfo
  {pages} {041023} (\bibinfo {year} {2017})}\BibitemShut {NoStop}%
\bibitem [{\citenamefont {Russo}\ \emph {et~al.}(2018)\citenamefont {Russo},
  \citenamefont {Barnes},\ and\ \citenamefont
  {Economou}}]{russo_photonic_2018}%
  \BibitemOpen
  \bibfield  {author} {\bibinfo {author} {\bibfnamefont {A.}~\bibnamefont
  {Russo}}, \bibinfo {author} {\bibfnamefont {E.}~\bibnamefont {Barnes}}, \
  and\ \bibinfo {author} {\bibfnamefont {S.~E.}\ \bibnamefont {Economou}},\
  }\href {\doibase 10.1103/PhysRevB.98.085303} {\bibfield  {journal} {\bibinfo
  {journal} {Physical Review B}\ }\textbf {\bibinfo {volume} {98}},\ \bibinfo
  {pages} {085303} (\bibinfo {year} {2018})}\BibitemShut {NoStop}%
\bibitem [{\citenamefont {Khaetskii}\ \emph {et~al.}(2002)\citenamefont
  {Khaetskii}, \citenamefont {Loss},\ and\ \citenamefont
  {Glazman}}]{khaetskii_electron_2002}%
  \BibitemOpen
  \bibfield  {author} {\bibinfo {author} {\bibfnamefont {A.~V.}\ \bibnamefont
  {Khaetskii}}, \bibinfo {author} {\bibfnamefont {D.}~\bibnamefont {Loss}}, \
  and\ \bibinfo {author} {\bibfnamefont {L.}~\bibnamefont {Glazman}},\ }\href
  {\doibase 10.1103/PhysRevLett.88.186802} {\bibfield  {journal} {\bibinfo
  {journal} {Physical Review Letters}\ }\textbf {\bibinfo {volume} {88}},\
  \bibinfo {pages} {186802} (\bibinfo {year} {2002})}\BibitemShut {NoStop}%
\bibitem [{\citenamefont {Khaetskii}\ \emph {et~al.}(2003)\citenamefont
  {Khaetskii}, \citenamefont {Loss},\ and\ \citenamefont
  {Glazman}}]{khaetskii_electron_2003}%
  \BibitemOpen
  \bibfield  {author} {\bibinfo {author} {\bibfnamefont {A.}~\bibnamefont
  {Khaetskii}}, \bibinfo {author} {\bibfnamefont {D.}~\bibnamefont {Loss}}, \
  and\ \bibinfo {author} {\bibfnamefont {L.}~\bibnamefont {Glazman}},\ }\href
  {\doibase 10.1103/PhysRevB.67.195329} {\bibfield  {journal} {\bibinfo
  {journal} {Physical Review B}\ }\textbf {\bibinfo {volume} {67}},\ \bibinfo
  {pages} {195329} (\bibinfo {year} {2003})}\BibitemShut {NoStop}%
\bibitem [{\citenamefont {Cywi\'nski}\ \emph {et~al.}(2009)\citenamefont
  {Cywi\'nski}, \citenamefont {Witzel},\ and\ \citenamefont
  {Sarma}}]{cywinski_pure_2009}%
  \BibitemOpen
  \bibfield  {author} {\bibinfo {author} {\bibfnamefont {L.}~\bibnamefont
  {Cywi\'nski}}, \bibinfo {author} {\bibfnamefont {W.~M.}\ \bibnamefont
  {Witzel}}, \ and\ \bibinfo {author} {\bibfnamefont {S.~D.}\ \bibnamefont
  {Sarma}},\ }\href@noop {} {\bibfield  {journal} {\bibinfo  {journal}
  {Physical Review B}\ }\textbf {\bibinfo {volume} {79}},\ \bibinfo {pages}
  {245314} (\bibinfo {year} {2009})}\BibitemShut {NoStop}%
\bibitem [{\citenamefont {Petrov}\ \emph {et~al.}(2009)\citenamefont {Petrov},
  \citenamefont {Kozlov}, \citenamefont {Ignatiev}, \citenamefont {Cherbunin},
  \citenamefont {Yakovlev},\ and\ \citenamefont {Bayer}}]{petrov_coupled_2009}%
  \BibitemOpen
  \bibfield  {author} {\bibinfo {author} {\bibfnamefont {M.~Y.}\ \bibnamefont
  {Petrov}}, \bibinfo {author} {\bibfnamefont {G.~G.}\ \bibnamefont {Kozlov}},
  \bibinfo {author} {\bibfnamefont {I.~V.}\ \bibnamefont {Ignatiev}}, \bibinfo
  {author} {\bibfnamefont {R.~V.}\ \bibnamefont {Cherbunin}}, \bibinfo {author}
  {\bibfnamefont {D.~R.}\ \bibnamefont {Yakovlev}}, \ and\ \bibinfo {author}
  {\bibfnamefont {M.}~\bibnamefont {Bayer}},\ }\href {\doibase
  10.1103/PhysRevB.80.125318} {\bibfield  {journal} {\bibinfo  {journal}
  {Physical Review B}\ }\textbf {\bibinfo {volume} {80}},\ \bibinfo {pages}
  {125318} (\bibinfo {year} {2009})}\BibitemShut {NoStop}%
\bibitem [{\citenamefont {Deng}\ and\ \citenamefont
  {Hu}(2005)}]{deng_nuclear_2005}%
  \BibitemOpen
  \bibfield  {author} {\bibinfo {author} {\bibfnamefont {C.}~\bibnamefont
  {Deng}}\ and\ \bibinfo {author} {\bibfnamefont {X.}~\bibnamefont {Hu}},\
  }\href {\doibase 10.1103/PhysRevB.72.165333} {\bibfield  {journal} {\bibinfo
  {journal} {Physical Review B}\ }\textbf {\bibinfo {volume} {72}},\ \bibinfo
  {pages} {165333} (\bibinfo {year} {2005})}\BibitemShut {NoStop}%
\bibitem [{\citenamefont {Gangloff}\ \emph {et~al.}(2018)\citenamefont
  {Gangloff}, \citenamefont {{\'Ethier-Majcher}}, \citenamefont {Lang},
  \citenamefont {Denning}, \citenamefont {Bodey}, \citenamefont {Jackson},
  \citenamefont {Clarke}, \citenamefont {Hugues}, \citenamefont {Gall},\ and\
  \citenamefont {Atat\"ure}}]{gangloff_quantum_2018}%
  \BibitemOpen
  \bibfield  {author} {\bibinfo {author} {\bibfnamefont {D.}~\bibnamefont
  {Gangloff}}, \bibinfo {author} {\bibfnamefont {G.}~\bibnamefont
  {{\'Ethier-Majcher}}}, \bibinfo {author} {\bibfnamefont {C.}~\bibnamefont
  {Lang}}, \bibinfo {author} {\bibfnamefont {E.}~\bibnamefont {Denning}},
  \bibinfo {author} {\bibfnamefont {J.}~\bibnamefont {Bodey}}, \bibinfo
  {author} {\bibfnamefont {D.}~\bibnamefont {Jackson}}, \bibinfo {author}
  {\bibfnamefont {E.}~\bibnamefont {Clarke}}, \bibinfo {author} {\bibfnamefont
  {M.}~\bibnamefont {Hugues}}, \bibinfo {author} {\bibfnamefont {C.~L.}\
  \bibnamefont {Gall}}, \ and\ \bibinfo {author} {\bibfnamefont
  {M.}~\bibnamefont {Atat\"ure}},\ }\href@noop {} {\bibfield  {journal}
  {\bibinfo  {journal} {arXiv:1812.07540}\ } (\bibinfo {year} {2018})},\
  \Eprint {http://arxiv.org/abs/1812.07540} {arXiv:1812.07540} \BibitemShut
  {NoStop}%
\bibitem [{\citenamefont {Botzem}\ \emph {et~al.}(2016)\citenamefont {Botzem},
  \citenamefont {McNeil}, \citenamefont {Mol}, \citenamefont {Schuh},
  \citenamefont {Bougeard},\ and\ \citenamefont
  {Bluhm}}]{botzem_quadrupolar_2016}%
  \BibitemOpen
  \bibfield  {author} {\bibinfo {author} {\bibfnamefont {T.}~\bibnamefont
  {Botzem}}, \bibinfo {author} {\bibfnamefont {R.~P.~G.}\ \bibnamefont
  {McNeil}}, \bibinfo {author} {\bibfnamefont {J.-M.}\ \bibnamefont {Mol}},
  \bibinfo {author} {\bibfnamefont {D.}~\bibnamefont {Schuh}}, \bibinfo
  {author} {\bibfnamefont {D.}~\bibnamefont {Bougeard}}, \ and\ \bibinfo
  {author} {\bibfnamefont {H.}~\bibnamefont {Bluhm}},\ }\href {\doibase
  10.1038/ncomms11170} {\bibfield  {journal} {\bibinfo  {journal} {Nature
  Communications}\ }\textbf {\bibinfo {volume} {7}},\ \bibinfo {pages} {11170}
  (\bibinfo {year} {2016})}\BibitemShut {NoStop}%
\bibitem [{\citenamefont {Korenev}(2007)}]{korenev_nuclear_2007}%
  \BibitemOpen
  \bibfield  {author} {\bibinfo {author} {\bibfnamefont {V.~L.}\ \bibnamefont
  {Korenev}},\ }\href {\doibase 10.1103/PhysRevLett.99.256405} {\bibfield
  {journal} {\bibinfo  {journal} {Physical Review Letters}\ }\textbf {\bibinfo
  {volume} {99}},\ \bibinfo {pages} {256405} (\bibinfo {year}
  {2007})}\BibitemShut {NoStop}%
\end{thebibliography}%
\end{document}